\documentclass[12pt,twoside,a4paper]{article}
\usepackage{graphicx}
\usepackage{url}
\usepackage{cite}
\usepackage{color}

\begin{document}
\thispagestyle{empty} 
\title{
\vskip-3cm
{\baselineskip14pt
\centerline{\normalsize DESY 18-036 \hfill ISSN 0418--9833}
\centerline{\normalsize MITP/18-020 \hfill} 
\centerline{\normalsize March 2018 \hfill}} 
\vskip1.5cm
\boldmath
{\bf $\Lambda_b^0$-baryon production in pp collisions in the } 
\\
{\bf general-mass variable-flavour-number scheme}
\\
{\bf and comparison with CMS and LHCb data}
\unboldmath
\author{
G.~Kramer$^1$ 
and H.~Spiesberger$^2$
\vspace{2mm} \\
\normalsize{
  $^1$ II. Institut f\"ur Theoretische
  Physik, Universit\"at Hamburg,
}\\ 
\normalsize{
  Luruper Chaussee 149, D-22761 Hamburg, Germany
} \vspace{2mm}\\
\normalsize{
  $^2$ Institut f\"ur Physik,
  Johannes-Gutenberg-Universit\"at,
}\\ 
\normalsize{
  Staudinger Weg 7, D-55099 Mainz, Germany,}\\
\normalsize{and Centre for Theoretical and
  Mathematical Physics and Department of Physics,}\\
\normalsize{
  University of Cape Town, Rondebosch 7700, South Africa}\\
\vspace{2mm} \\}}
%
\date{}
\maketitle
\begin{abstract}
\medskip
\noindent
We calculate the next-to-leading-order cross section for the 
inclusive production of $\Lambda_b$ baryons in $pp$ collisions 
in the general-mass variable-flavor-number scheme. We use 
realistic evolved non-perturbative fragmentation functions 
obtained from fits to $B$-meson production in $e^+e^-$ 
annihilation and compare our results for transverse-momentum 
and rapidity distributions with recent experimental data from 
the CMS and the LHCb collaborations at the CERN LHC. We find 
satisfactory agreement in general, with some indication for 
the need to modify the available fragmentation functions at 
larger values of the scale variable. 
\\
\\
PACS: 12.38.Bx, 12.39.St, 13.85.Ni, 14.20.Mr
\end{abstract}
\clearpage


\section{Introduction}

The study of the inclusive production of hadrons containing $b$ 
quarks plays a particularly important role in testing quantum 
chromodynamics (QCD). The predictions in the framework of 
perturbative QCD are based on the factorization approach. 
Cross sections are calculated as a convolution of three terms: 
the parton distribution functions (PDF) encoding the parton 
content of the initial hadronic state, the partonic hard 
scattering cross sections computed as a perturbative series 
in powers of the strong coupling constant, and the 
fragmentation functions (FF), which describe the production 
yield and the momentum distribution for a given $b$ hadron 
in a parton. 

In the past, measurements of inclusive $b$-hadron production 
and the corresponding QCD calculations have been done mostly 
for $B$ mesons, i.e., $B^\pm$, $B^0$, $\bar{B}^0$, $B_s^0$, 
and $\bar{B}_s^0$. Data for $p\bar{p}$ collisions at $\sqrt{S} 
= 1.96$~TeV have been obtained at the FNAL Tevatron 
Collider~\cite{Acosta:2004yw,Abulencia:2006ps} and for $pp$ 
collisions at $\sqrt{S} = 7$, 8 and 13~TeV at the CERN Large 
Hadron Collider (LHC) by the CMS, ATLAS and LHCb 
collaborations~\cite{Khachatryan:2011mk,Chatrchyan:2011pw, 
Chatrchyan:2011vh,ATLAS:2013cia,Aaij:2012jd,Khachatryan:2016csy}. 
The first measurement of the production cross section of a 
$b$ baryon, $\Lambda_b^0$, has been performed by the CMS 
collaboration at the LHC~\cite{Chatrchyan:2012xg} at $\sqrt{S} 
= 7$~TeV using fully reconstructed $\Lambda_b^0 \to J/\psi 
\Lambda$ decays. CMS has measured the cross section as a 
function of the transverse momentum $p_T$ and the rapidity 
$y$ of the produced $\Lambda_b^0$ in the region $10 \leq p_T 
\leq 50$~GeV and in the central rapidity region $0 \leq |y| 
\leq 2$. Also the cross section ratio $\sigma(\bar{\Lambda}_b) 
/ \sigma(\Lambda_b)$ has been obtained in the same kinematic 
range. Later, the LHCb collaboration has published similar 
measurements in the forward rapidity region $2.0 \leq y \leq 
4.5$ in the $p_T$ range $0 < p_T < 20$~GeV for $\sqrt{S} = 7$ 
and $8$~TeV \cite{Aaij:2015fea}. Here, the measurement of 
$\Lambda_b^0$ production was based on the observation of the 
decay $\Lambda_b^0 \rightarrow J/\psi p K^-$.

Inclusive production of $\Lambda_b$ baryons is of interest 
for several reasons. First, there is the question whether the 
perturbative approach to calculate $b$-quark production cross 
sections is likewise applicable for the production of $b$ 
baryons as it is for the production of $B$ mesons. Second, 
there is the more important question about details of the 
fragmentation of $b$ quarks and other partons, as for example 
of gluons, into $b$ baryons. So far there exists almost no 
information on the $\Lambda_b$ FF from earlier experiments. 
The new data from experiments at the LHC are therefore first 
of all valuable as they provide us with information needed 
to determine the $\Lambda_b$ FF. The comparison of data for 
the production of $\Lambda_b$ baryons with data for $B$-meson 
production could reveal unexpected differences between the 
$\Lambda_b$-baryon and $B$-meson fragmentation functions. 

Third, there is the problem that incompatible results for the 
$b$ hadron production fractions have been found in different 
measurements. The relative production rates of $b$ hadrons are 
described by the fragmentation fractions $f_u$, $f_d$, $f_s$, 
$f_c$, and $f_{\rm baryon}$ for the probability that a $b$ 
quark fragments into a $B_q$ meson ($q = u, d, s, c$) or a 
$b$ baryon. It is assumed that $f_u = f_d$, and $f_{\rm baryon} 
= f_{\Lambda_b}$ is obtained from $\Lambda_b$ production. 
According to the most recent analysis of the Heavy Flavor 
Averaging Group (HFAG~\cite{Amhis:2016xyh}) one finds  $f_u 
= f_d = 0.412 \pm 0.008$, $f_s = 0.088 \pm 0.013$, and 
$f_{\rm baryon} = 0.089 \pm 0.012$ when determined from LEP 
data for $Z \rightarrow b\bar{b}$ decays only, but Tevatron 
data lead to $f_u = f_d = 0.340 \pm 0.021$, $f_s = 0.101 \pm 
0.015$, and $f_{\rm baryon} = 0.218 \pm 0.047$. Only for 
$f_s$ these values agree well, but for the $\Lambda_b$ baryon 
fragmentation fraction there is a discrepancy of more than a 
factor of two. These results are clearly not compatible with 
the assumption that the $b$-hadronization fractions are 
universal. Further evidence for this non-universality came 
from LHCb data where a strong dependence of $f_{\Lambda_b}$ 
on the transverse momentum was observed~\cite{Aaij:2011jp, 
Aaij:2014jyk}. The discrepancy of results for $f_{\Lambda_b}/f_d$ 
measured at LEP or at hadron colliders may indicate a strong 
dependence on the kinematic properties of the produced $b$ 
quark, as suggested for example in~\cite{Aaltonen:2008zd}. 
The $b$ jets in $Z$ decays at LEP have $p_T \simeq 40$ GeV 
while the average $p_T$ of the measurement at CDF is 10~GeV. 
Measurements at the LHCb experiment probe an even lower $p_T$ 
range. Such a strong scale dependence is, however, barely 
consistent with the theoretical predictions. For example, in 
Ref.~\cite{Kniehl:2008zza} we could demonstrate that the 
$b \to B$ fragmentation fraction (denoted $B(\mu)$ in 
\cite{Kniehl:2008zza} and evaluated as the integral over 
the $b \to B$ FF) depends only very little on the scale $\mu$ 
varied in the range between 4.5 and 91.2~GeV. If the difference 
of data for $f_{\Lambda_b}/f_d$ obtained from LEP or from 
$p\bar{p}$ $(pp)$ colliders are confirmed, one should conclude 
that the production mechanism for $\Lambda_b$ (and alike for 
other $b$ hadrons) is affected by the presence of strongly 
interacting particles in the initial state, e.g., by the 
proton remnants emitted in the extreme forward direction. 

In our calculation of $\Lambda_b$-baryon production we shall 
use the FF for $b \rightarrow B$ as obtained in 
Ref.~\cite{Kniehl:2008zza} from LEP data~\cite{Heister:2001jg,
Abbiendi:2002vt,Abe:1999ki,Abe:2002iq}. For the light-quark 
fragmentation fractions we shall assume in the following the 
value $f_u = f_d = 0.401$ \cite{Beringer:1900zz}, which is 
very close to the value in our previous work \cite{Kniehl:2008zza}.

It is the purpose of this work to study the cross section for 
inclusive production of $\Lambda_b$ baryons in the framework 
of the general-mass variable-flavour-number scheme (GM-VFNS) 
\cite{Kniehl:2004fy,Kniehl:2005mk}. This framework has provided 
a good description for $b$-meson production in $p\bar{p}$ 
collisions at $\sqrt{S}$ = 1.96 TeV at the FNAL Tevatron 
Collider~\cite{Kniehl:2008zza} and in $pp$ collisions at 
$\sqrt{S} = 7$~TeV at the CERN LHC by the CMS, ATLAS and 
LHCb collaborations~\cite{Kniehl:2011bk,Kniehl:2015fla,
Khachatryan:2011mk,Chatrchyan:2011pw,Chatrchyan:2011vh,
ATLAS:2013cia,Aaij:2012jd}.

The GM-VFNS is essentially the conventional next-to-leading order 
(NLO) QCD parton-model approach supplemented with finite-mass 
effects, intended to improve the description at small and medium 
transverse momenta $p_T$. The original GM-VFNS formulation 
\cite{Kniehl:2004fy,Kniehl:2005mk,Kniehl:2012ti} was, however, 
not suitable for the calculation of the cross section at very 
small transverse momenta $p_T$. This was caused by the specific 
choice of the scale parameter $\mu_I$ for the initial-state 
factorization as $\mu_I = \sqrt{m_b^2 + p_T^2}$, where $m_b$ 
is the mass of the $b$ quark. As a consequence of this choice, 
only at $p_T = 0$ does the scale parameter approach $\mu_I = 
m_b$ where the $b$-quark parton distribution function (PDF) 
vanishes (in almost all available PDF parametrizations). 
Therefore the transition to the fixed-flavour number scheme 
(FFNS), which is the appropriate scheme for calculating 
$d\sigma/dp_T$ at small $p_T < m_b$, is not reached for non-zero 
$p_T > 0$. The original GM-VFNS prescription was therefore 
modified later. In Ref.~\cite{Kniehl:2015fla,Kramer:2015wda} 
we have shown that a smooth transition to the FFNS at finite 
$p_T > 0$ can be obtained by choosing the factorization scale 
appropriately. With the choice $\mu_I = 0.5 \sqrt{m_b^2+p_T^2}$ 
instead of $\mu_I = \sqrt{m_b^2+p_T^2}$ a reasonably good 
description of the experimental data for $B$-meson production 
down to $p_T = 0$ could be achieved for the CDF data 
\cite{Acosta:2004yw} in $p\bar{p}$ collision at the Tevatron 
and for the LHCb data \cite{Aaij:2013noa} in $pp$ collisions 
at the LHC at $\sqrt{S} = 7$~TeV. Since the recent measurements 
of the inclusive $\Lambda_b$ production cross sections at the 
LHCb extend down to $p_T = 0$ \cite{Aaij:2015fea} we will apply 
the GM-VFNS with this modified scale choice. For the CMS data 
which are at higher $p_T$ well above the $b$-quark threshold, 
we keep the original setting of scales. 

The plan of the paper is as follows. In Section 2 we introduce 
our strategy and describe our choice of the proton PDFs and 
the FFs for the transition $b \rightarrow \Lambda_b^0$. In 
Section 3 we collect our results for inclusive $\Lambda_b$ 
production at $\sqrt{S} = 7$~TeV and compare with the CMS 
data published in Ref.~\cite{Chatrchyan:2012xg}. A similar 
comparison is then performed in Section 4 for LHCb 
data~\cite{Aaij:2015fea} at $\sqrt{S} = 7$ and 8~TeV in the
forward rapidity range $2.0 < y < 4.5$. Here we study also 
the cross section ratios of the 7 and 8~TeV data and ratios 
of the inclusive production cross sections for $\Lambda_b$ 
and $B^0$ mesons. Our conclusions are presented in Section 5.


\section {Setup and Input}

The theoretical foundation of the GM-VFNS framework as well 
as technical details of its implementation have been presented 
previously in Refs.~\cite{Kniehl:2004fy,Kniehl:2005mk}. 
Here we describe only the input required for the numerical 
calculations discussed below. For the proton PDFs we use 
the set CTEQ14~\cite{Dulat:2015mca} as implemented in the 
LHAPDF library~\cite{Buckley:2014ana}. We take the $b$-quark 
pole mass to be $m_b = 4.5$~GeV and evaluate the strong 
coupling $\alpha_s^{(n_f)}(\mu_R)$ at NLO with  
$\Lambda_{\overline{MS}}^{(4)} = 328$~MeV for $n_f$ = 4. 
This corresponds to $\Lambda_{\overline{MS}}^{(5)} = 226$~MeV 
above the 5-flavor threshold of the renormalization scale 
chosen at $\mu_R = m_b$. 

For simplicity, in the following sections we take the initial- 
and final-state factorization scales, entering the PDFs and FFs, 
respectively, to have the same value, denoted by $\mu_F$. We 
choose $\mu_F$ and the renormalization scale $\mu_R$, at which 
$\alpha_s$ is evaluated, to be $\mu_F = \xi_F \mu_0$ and $\mu_R 
= \xi_R \mu_0$, where $\mu_0$  will be specified in the next 
two sections, when we compare our results to the CMS and LHCb 
data. In the calculation of cross sections to be compared with 
the CMS data we shall vary the parameters $\xi_F$ and $\xi_R$ 
about their default values $\xi_F = \xi_R = 1$ up and down by 
factors of 2 (with the restriction $1/2 < \xi_R/\xi_F < 2$). 
For comparisons with the LHCb data we restrict ourselves to 
variations of the renormalization scale factor $\xi_R$.

We employ the non-perturbative $B$-meson FFs determined in 
\cite{Kniehl:2008zza}. These FFs were obtained by fitting 
experimental data for inclusive production in $e^+e^-$ 
annihilation taken by the ALEPH \cite{Heister:2001jg}, and OPAL 
collaborations~\cite{Abbiendi:2002vt} at CERN LEP1 and by the 
SLD collaboration~\cite{Abe:1999ki,Abe:2002iq} at SLAC SLC. 
Since these data were all taken on the $Z$-boson resonance, 
$\alpha_s^{(n_f)}(\mu_R)$ was evaluated with $n_f = 5$ and 
the renormalization and factorization scales were fixed at 
$\mu_F = \mu_R = m_Z$ in Ref.~\cite{Kniehl:2008zza}. The 
starting scale $\mu_0$ of the $b \rightarrow B$ FF was chosen 
to be $\mu_0 = m_b$ in accordance with Ref.~\cite{Dulat:2015mca}. 
Below $\mu_F = \mu_0$ the light-quark and gluon FFs for $q,g 
\rightarrow B$ (including the charm quark, i.e., $q = u, d, 
s, c$) were assumed to vanish. A simple power ansatz has 
yielded the best fit to the experimental data. 

One should notice that the $B$-meson FFs of 
Ref.~\cite{Kniehl:2008zza} do not distinguish between different 
$b$-hadron states. In fact, the OPAL~\cite{Abbiendi:2002vt} 
and SLD~\cite{Abe:1999ki,Abe:2002iq} data include all $b$ 
hadrons, i.e., the mesons $B^{\pm}$, $B^0/\bar{B}^0$ and 
$B_s^0/\bar{B}_s^0$ as well as $b$-baryons, such as 
$\Lambda_b^0$, while in the ALEPH~\cite{Heister:2001jg} 
analysis, only final states with identified $B^{\pm}$ and 
$B^0/\bar{B}^0$ mesons were taken into account\footnote{
  A more recent study of $b$-hadron fragmentation 
  by the DELPHI collaboration~\cite{DELPHI:2011aa} is using 
  $b$-tagged final states which also combines all $b$-hadrons. 
  These data agree with the earlier ALEPH, OPAL and SLD 
  measurements within their experimental uncertainties.
}. 
Despite of the differences in the experimental analyses of 
ALEPH, OPAL and SLD it was assumed in Ref.~\cite{Kniehl:2008zza} 
that the data can be described by one common FF. They were 
normalized to provide cross sections for the $B^+$ or $B^0$ 
mesons. They can also be used to calculate $b$-quark 
production by removing the fragmentation fraction for the 
$b \to B^+$ transition which was assumed as $f_d = f_u 
= 0.397$ in Ref.~\cite{Kniehl:2008zza}. We assume that 
also $\Lambda_b$ production is described by the same FF 
and that only the normalization has to be adjusted, i.e., 
the same FF multiplied by $f_{\Lambda_b}/f_d$ can be used 
to obtain the FF for $b \rightarrow \Lambda_b^0$.


\section {
Comparison with CMS data}

CMS has measured the cross section for $\Lambda_b^0$ production 
at $\sqrt{S}$ = 7 TeV in the central rapidity region $-0.2 < y 
< 2.0$ and for $p_T$ in the range $10 < p_T < 50$~GeV  
\cite{Chatrchyan:2012xg}. The data are given for the differential 
cross sections $d\sigma/dp_T$ and $d\sigma/d|y|$ multiplied with 
the branching ratio for the decay $\Lambda_b^0 \rightarrow 
J/\psi \Lambda$. We calculate $d\sigma/dp_T$ integrated 
over $|y|$ and $d\sigma/dy$ integrated over the considered
$p_T$ range with the same binning as chosen by CMS.

The default renormalization and factorization scale is chosen 
as $\mu_0 = \sqrt{m_b^2+p_T^2}$ and an estimate of theoretical 
uncertainties is obtained by varying $\xi_R$ and $\xi_F$ as 
described above. We multiply the calculated cross sections 
by $Br(\Lambda_b^0 \to J/\psi \Lambda) f_{\Lambda_b}/f_d 
= (1.45 \pm 0.20) \times 10^{-4}$. This is obtained from 
$f_d = 0.401$ and the latest PDG value for $f_{\Lambda_b} 
Br(\Lambda_b^0 \to J/\psi \Lambda) = (5.8 \pm 0.8) \times 10^{-5}$ \cite{Patrignani:2016xqp}.

\begin{figure}[t]
\begin{center}
\includegraphics[width=7.8cm]{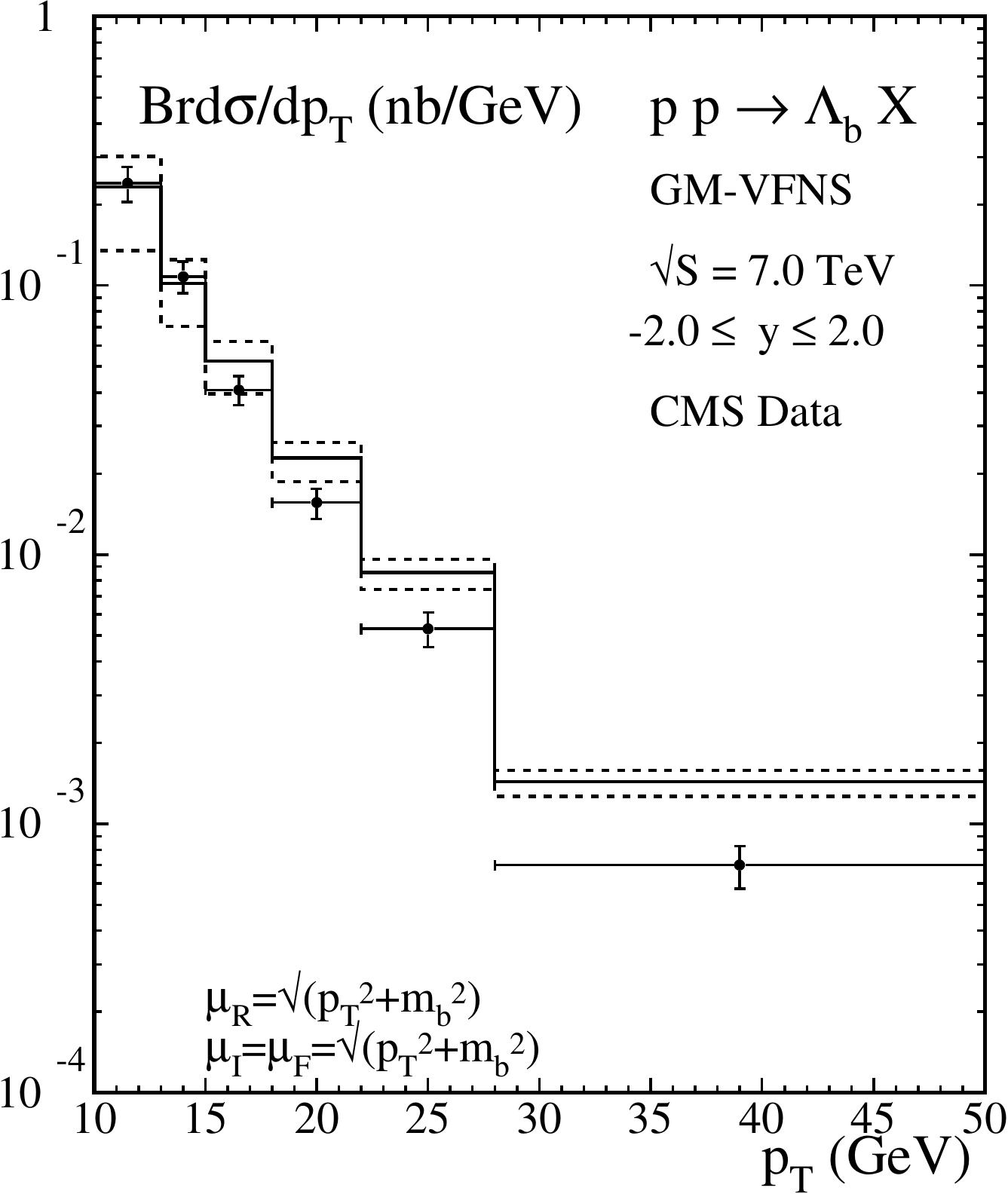}
\raisebox{-1.4mm}{\includegraphics[width=7.8cm]{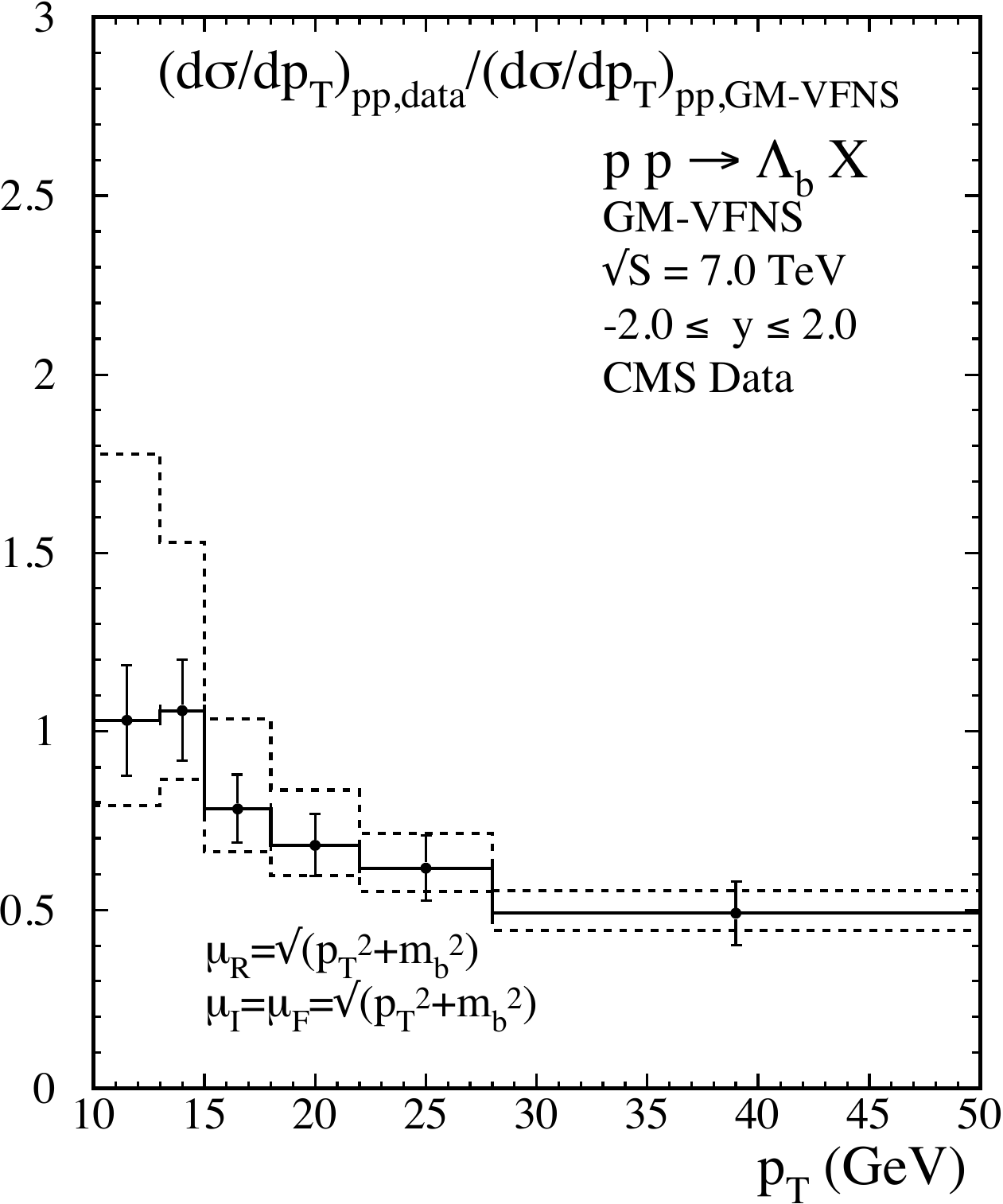}}
\end{center}
\caption{
  \label{fig:1} 
  Left panel: Differential cross section $d\sigma/dp_T$ times 
  branching ratio $Br(\Lambda_b^0 \to J/\psi\Lambda)$ of prompt 
  inclusive $\Lambda_b^0$-baryon production in the GM-VFNS for 
  $\sqrt{S} = 7.0$~TeV $pp$ collisions with $|y| < 2.0$ compared 
  to CMS data~\cite{Chatrchyan:2012xg}. The upper and lower dashed 
  histograms are calculated with $\mu_R$ and $\mu_I =\mu_F$ changed 
  independently by factors of $1/2$ and $2$ with the restriction 
  $1/2 < \mu_R/\mu_F < 2$. 
  Right panel: Ratios of data over theory. For the central, 
  full-line histogram the CMS data are normalized to the 
  calculated cross sections for the default scale (full line 
  in the left panel). The upper/lower dashed-line histograms 
  are the ratios of data normalized to the predictions with 
  scales that lead to the minimal/maximal cross sections  
  (dashed lines in the left panel). Experimental uncertainties 
  are shown by error bars only for the central curve. 
  }
\end{figure}

\begin{figure}[tb]
\begin{center}
\includegraphics[width=7.8cm]{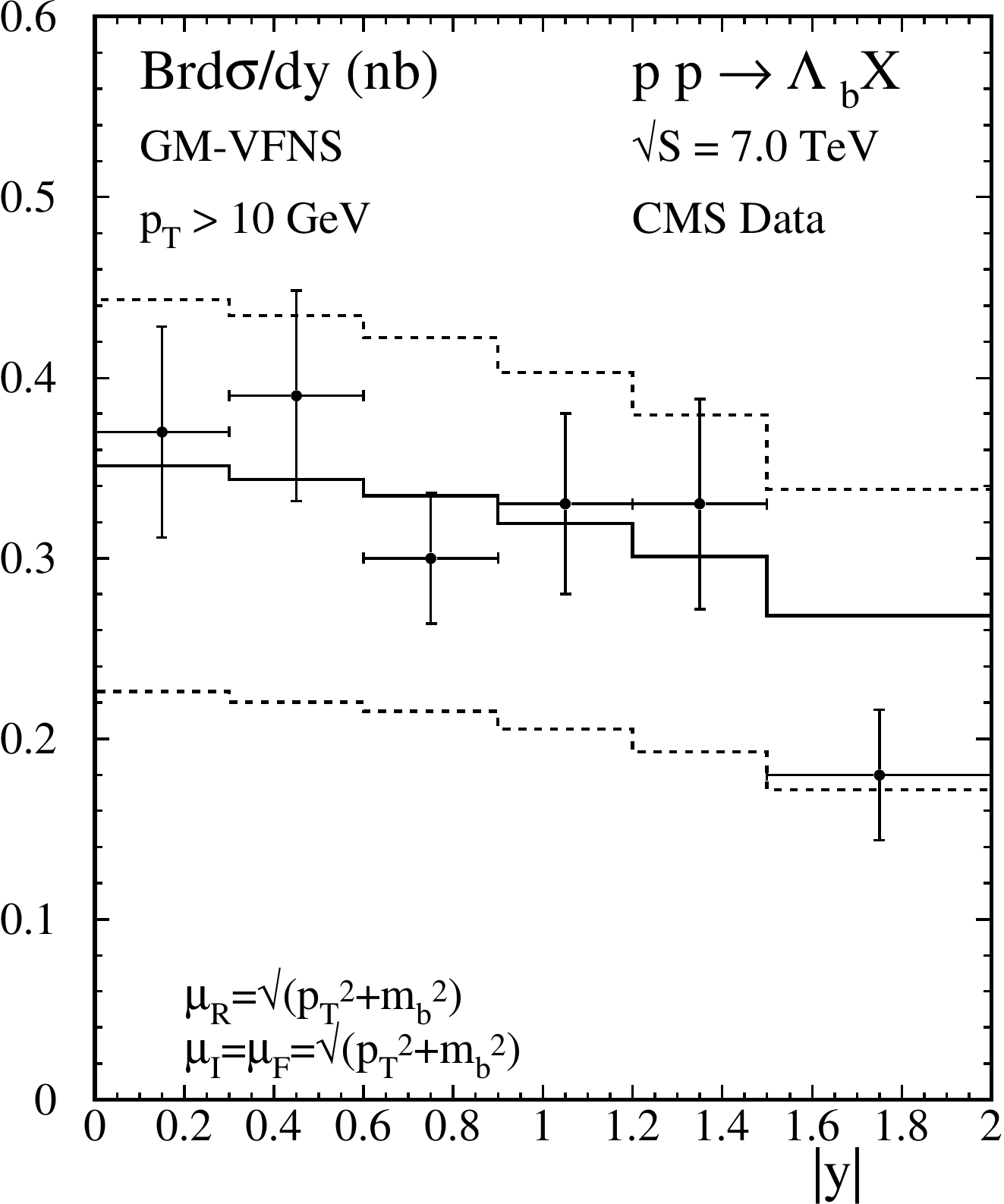}
\raisebox{-0.6mm}{\includegraphics[width=7.8cm]{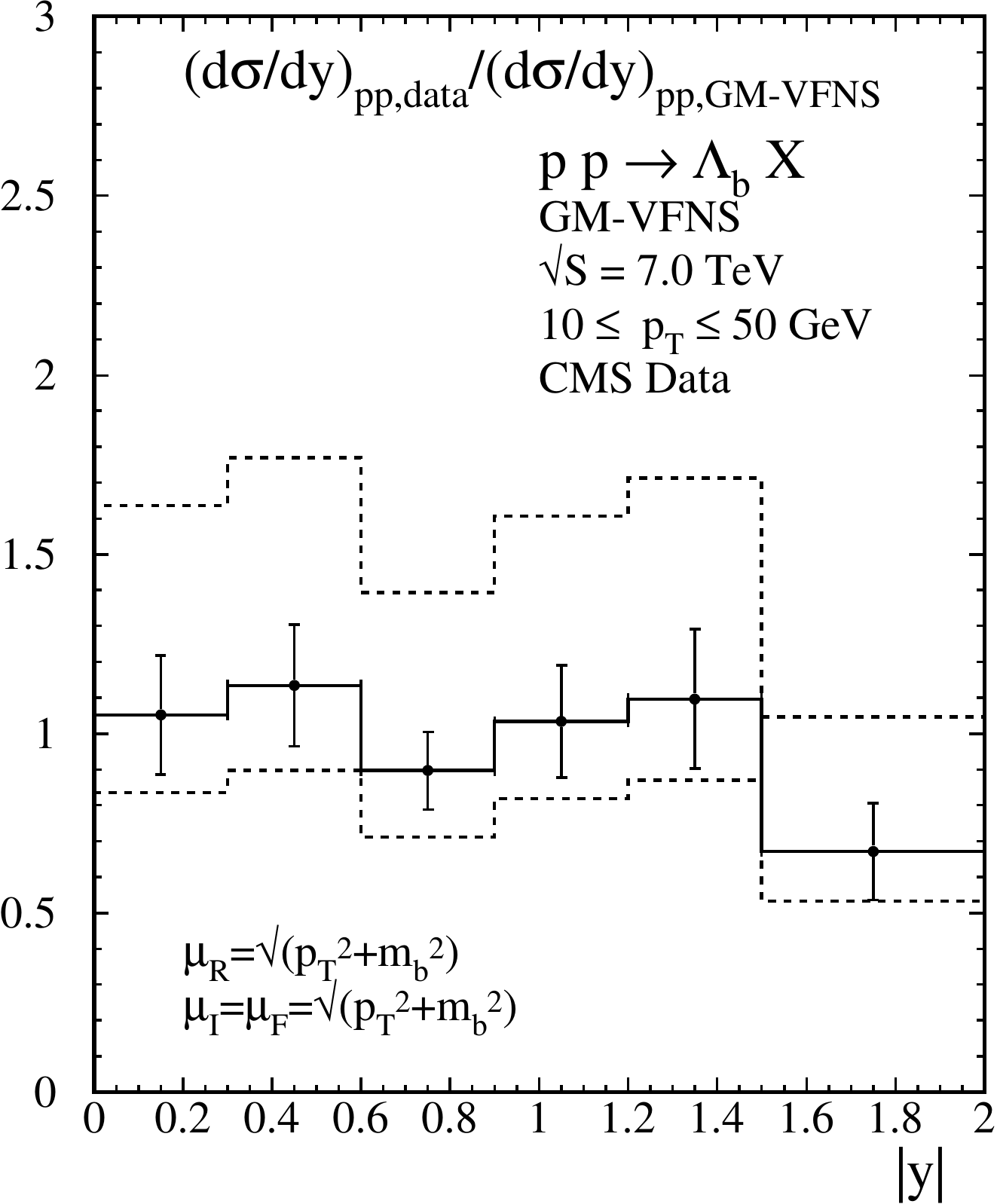}}
\end{center}
\caption{
  \label{fig:2} 
  Differential cross section $d\sigma/dy$ times branching ratio 
  for prompt inclusive $\Lambda_b^0$-baryon production in the 
  GM-VFNS for $\sqrt{S} = 7.0$~TeV $pp$ collisions for $10 \leq 
  p_T \leq 50$~GeV compared with CMS data~\cite{Chatrchyan:2012xg}. 
  The upper and lower dashed histograms are calculated with 
  $\mu_R$ and $\mu_I = \mu_F$ varied independently by factors of 
  $1/2$ and $2$. The ratios of data over theory in the right 
  panel are calculated as described above (see caption of 
  Fig.~\ref{fig:1}).
  }
\end{figure}

Our results are compared with the experimental data in 
Figs.~\ref{fig:1} and \ref{fig:2}. The errors of the experimental 
data points are obtained from Ref.~\cite{Chatrchyan:2012xg} by 
adding in quadrature the statistic and systematic errors quoted 
there. The error of the branching ratio $Br(\Lambda_b^0 \to 
J/\psi\Lambda)$ does not enter. The left side of Fig.~\ref{fig:1} 
shows $d\sigma/dp_T$ times branching ratio integrated over the 
rapidity $|y| < 2.0$ and the left side of Fig.~\ref{fig:2} shows 
$d\sigma/dy$ integrated over $p_T$ in the measured range $10 < 
p_T < 50$~GeV. We find agreement between theory and data for the 
three lowest $p_T$ bins within the theory error band, but for the 
three upper $p_T$-bins the experimental values of $d\sigma/dp_T$ 
lie outside the error band from variations of the scale parameters. 
The prediction of $d\sigma/dp_T$ for the largest $p_T$ bin is 
larger by a factor of approximately 2 as compared to the 
experimental point. This is seen more clearly in the right 
panel of Fig.~\ref{fig:1} where we have plotted the ratio of 
the measured cross section with respect to the theory prediction. 
The full-line histogram is obtained from data normalized to the 
prediction with the default scale, the dashed-line histograms 
are found when normalizing data to the minimal scale choice 
(upper dashed-line histogram) or maximal scale choice (lower 
dashed-line histogram). For readability the experimental errors 
are shown only for the default scale. The comparison in 
Fig.~\ref{fig:1} may be taken as an indication that the FF 
for $b \rightarrow \Lambda_b$ behaves differently compared 
to the FF for $b \to B$ meson. For example, the stronger 
decrease of the data with increasing $p_T$ could be obtained 
if the maximum of the $\Lambda_b$ FF was shifted to smaller 
$z$ as compared to the $B$-meson FF. A similar conclusion was 
suggested in Ref.~\cite{Chatrchyan:2012xg}, based on a comparison 
of data for $\Lambda_b^0$, $B^+$, and $B^0$ production with 
predictions from POWHEG~\cite{Alioli:2010xd,Frixione:2007nw} 
and PYTHIA~\cite{Sjostrand:2006za}. The predictions from PYTHIA 
agree with data at low $p_T$, but over-estimate data at large 
$p_T$, very similar to our results. In contrast, POWHEG 
is in better agreement with data for the large-$p_T$ bins, 
but below data at low $p_T$. Since low $p_T$ dominate for 
all values of $y$, the POWHEG prediction for $d\sigma/dy$ 
is below data by almost a factor of 2 in the whole 
$y$ range. 

The prediction for $d\sigma/dy$ as a function of $|y|$ between 
0 and 2 is shown in Fig.~\ref{fig:2}, left panel, compared to 
the CMS data. It agrees quite well with the experimental data 
even for the default scale choice, except for the bin $1.5 < 
|y| < 2.0$. This is, of course, consistent with the comparison 
of the $p_T$-differential cross section. The ratio of data and 
the calculated cross sections times branching ratio is shown in 
the right panel of Fig.~\ref{fig:2}.


\section {
Comparison with LHCb data}

\begin{figure}[tb]
\begin{center}
\includegraphics[width=7.8cm]{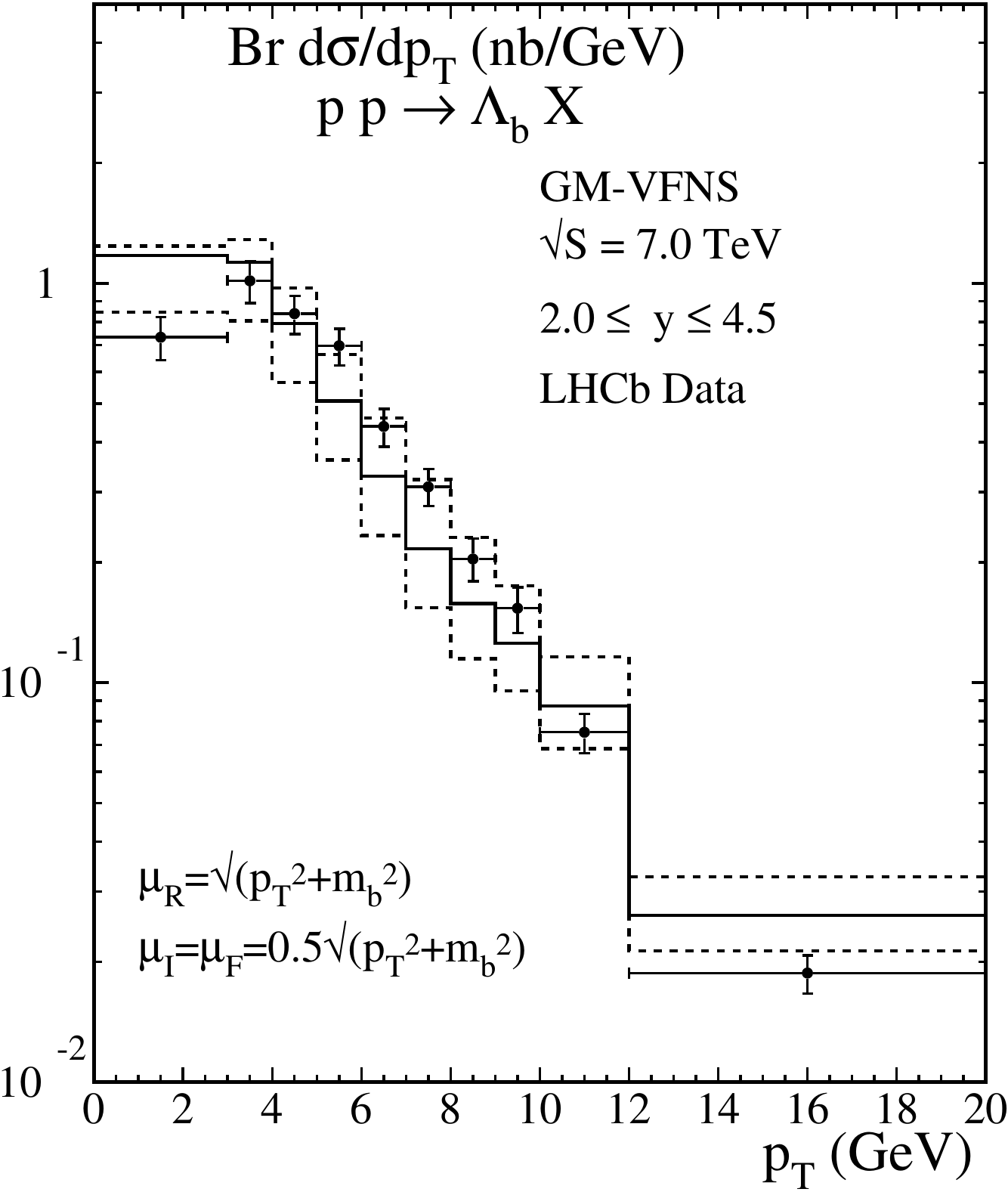}
\raisebox{-0.6mm}{\includegraphics[width=7.8cm]{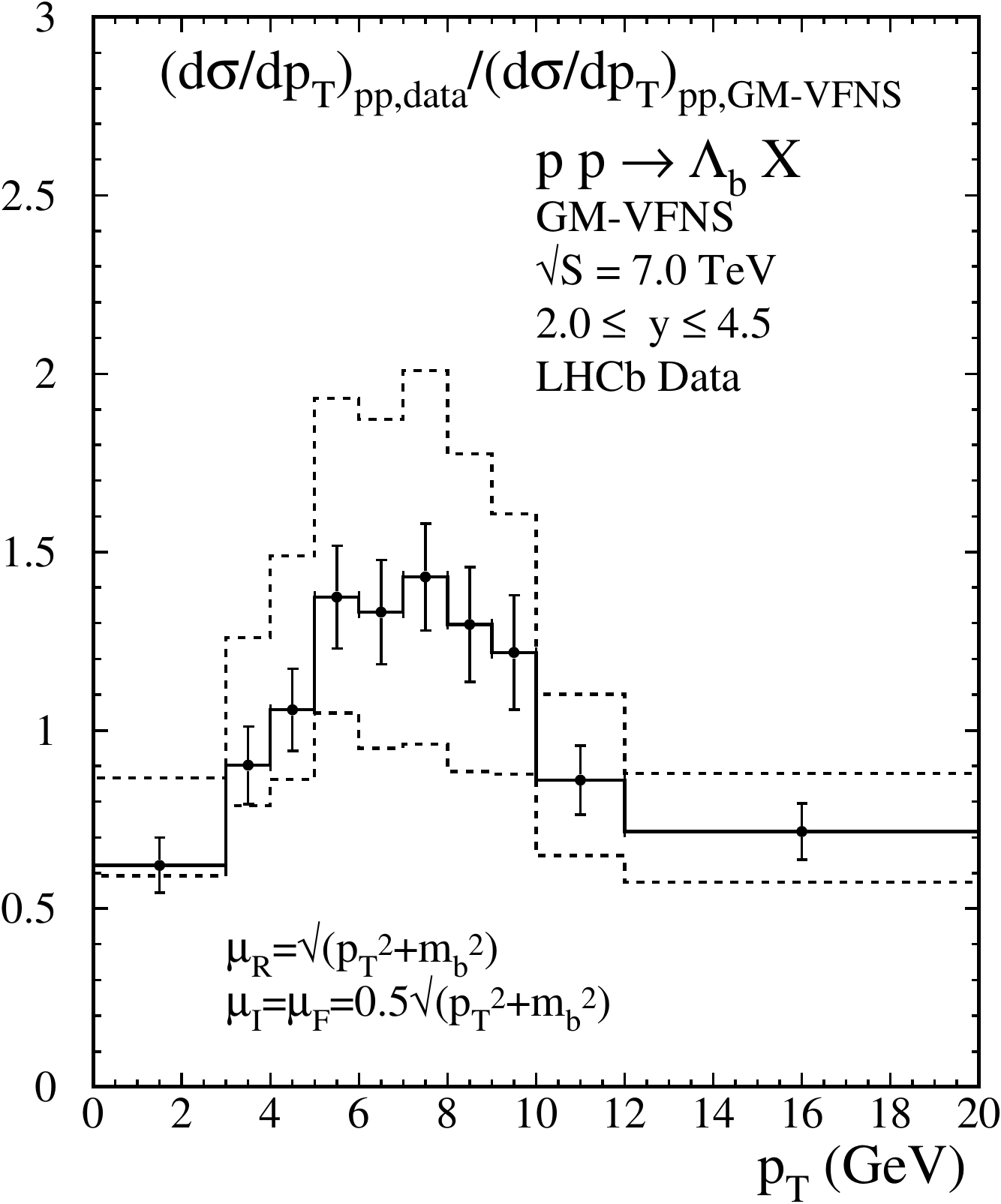}}
\end{center}
\caption{
  \label{fig:3} 
  Differential cross section $d\sigma/dp_T$ times branching 
  ratio for prompt inclusive $\Lambda_b^0$-baryon production 
  in the GM-VFNS for $\sqrt{S} = 7.0$~TeV $pp$ collisions with 
  $2.0 \leq y \leq 4.5$ compared to LHCb data~\cite{Aaij:2015fea}. 
  The upper and lower dashed histograms are calculated with 
  $\mu_R$ changed by factors $1/2$ and $2$. The ratios of data 
  over theory in the right panel are calculated as described 
  above (see caption of Fig.~\ref{fig:1}).
  }
\end{figure}

The LHCb collaboration has measured $\Lambda_b^0$ production in 
$pp$ collisions at $\sqrt{S} = 7$~TeV in 2011 and $\sqrt{S} = 
8$~TeV in 2012 \cite{Aaij:2015fea}. The data extend to small 
transverse momenta, $0 < p_T < 20$~GeV, in the forward-rapidity 
range, $2.0 < y < 4.5$. The $p_T$-differential cross sections 
are presented in five rapidity bins between $y = 2.0$ and 4.5 
and in 10 (12) $p_T$ bins between 0 and 20~GeV for the 7~TeV 
(8~TeV) measurements, respectively. Since the shape of the 
distributions as a function of $p_T$ is very similar for all 
values of $y$ and differ only by their normalization, we find 
it sufficient to compare our predictions with data for 
$d\sigma/dp_T$ integrated over the full $y$ range, $2.0 < y 
< 4.5$. We determine the corresponding values by summing the 
original data given in Ref.~\cite{Aaij:2015fea} over the five 
$y$ bins. The uncertainties are obtained correspondingly by 
adding the statistical and systematic errors of the bins 
linearly and then combining statistical and systematic errors 
in quadrature to obtain a total uncertainty. 

The $\Lambda_b^0$ baryons were identified in the decay 
$\Lambda_b^0 \rightarrow J/\psi p K^-$ and results are therefore 
given as cross sections times branching ratio of this decay. 
This branching ratio is deduced in \cite{Aaij:2015fea} from 
the ratio of the cross sections for $\Lambda_b^0$ and $\bar{B}^0$ 
production at 7 and 8~TeV. Unfortunately the cross section data 
in Ref.~\cite{Aaij:2015fea} are given for the sum of $B^0 + 
\bar{B}^0$ production and not just for $\bar{B}^0$ production. 
Therefore the ratio shown in Fig.~6 of \cite{Aaij:2015fea} 
should be multiplied by two to determine $R_{\Lambda_b^0 / 
\bar{B}^0}$ and we find for the branching ratio $Br(\Lambda_b^0 
\rightarrow J/\psi p K^-) = (6.34 \pm 1.24) \times 10^{-4}$ 
where the errors given in \cite{Aaij:2015fea} are summed in 
quadrature. The analysis of \cite{Aaij:2015fea} leading to 
this value is based on input for $f_{\Lambda_b}/f_d$ 
which was taken from Fig.~3 (right panel) of 
Ref.~\cite{Aaij:2014jyk}. At $p_T$ = 5 GeV one has 
$f_{\Lambda_b}/f_d = 0.50$. This is combined in 
$Br(\Lambda_b^0 \rightarrow J/\psi p K^-) f_{\Lambda_b}/f_d 
= (3.17 \pm 0.62) \times 10^{-4}$. We shall use this value 
for the branching ratio times $f_{\Lambda_b}/f_d$ in our 
calculations to obtain the cross section for $\Lambda_b^0$ 
production. 

In the low $p_T$ range relevant for the LHCb data we have 
to choose the factorization scale following our previous 
work \cite{Kniehl:2015fla,Kramer:2015wda}. Only with 
$\mu_I = \mu_F = 0.5 \sqrt{m_b^2 + p_T^2}$ (instead of 
$\mu_I = \mu_F = \sqrt{m_b^2 + p_T^2}$ as in the previous 
section) we find a smooth transition of the GM-VFNS 
prescription to the FFNS. The default renormalization scale 
is fixed at $\mu_R = \sqrt{p_T^2+m_b^2}$ and variations by 
factors of 1/2 and 2 are studied to obtain an estimate of 
the theoretical uncertainty. 

\begin{figure}[tb]
\begin{center}
\includegraphics[width=7.8cm]{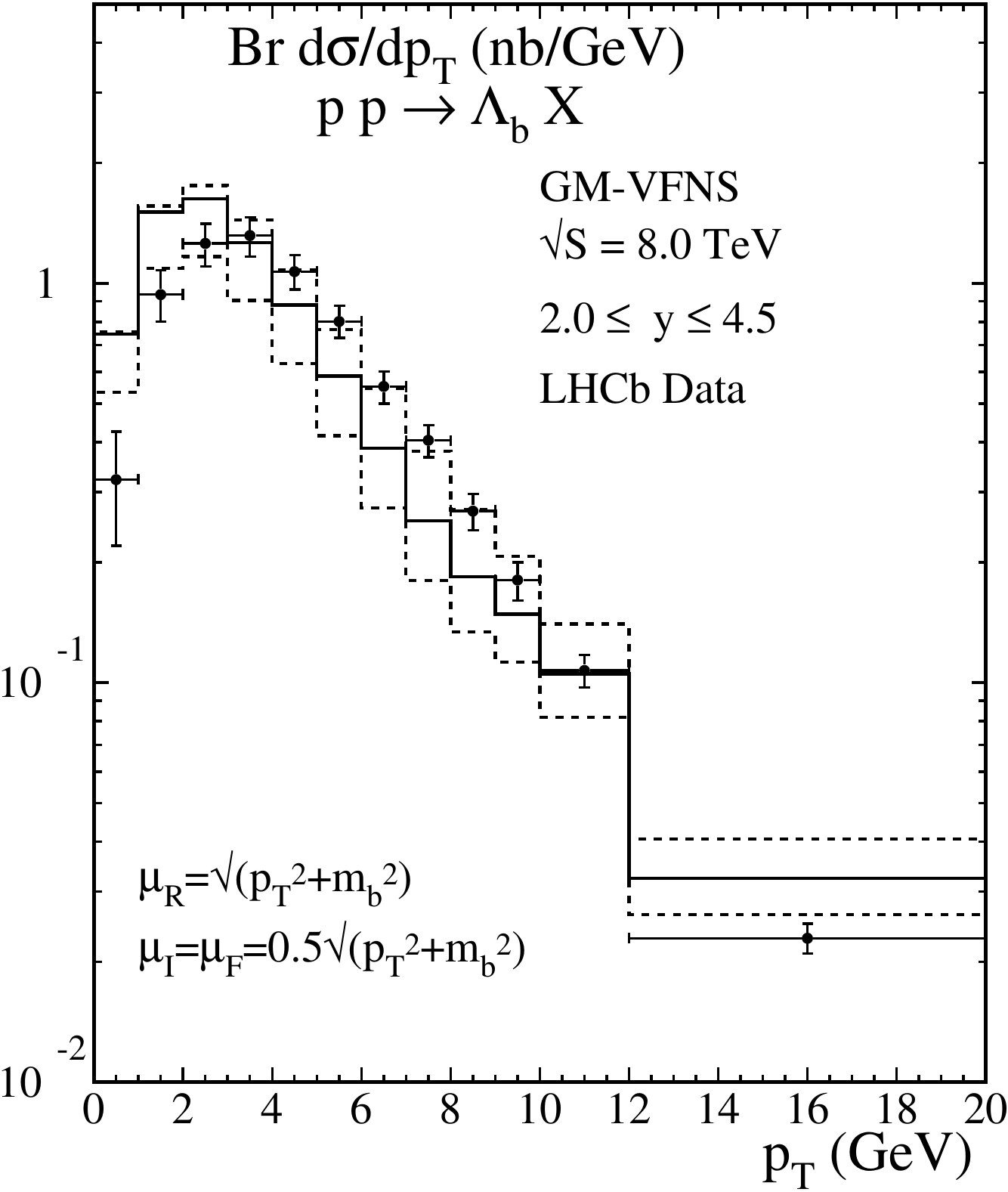}
\raisebox{-0.6mm}{\includegraphics[width=7.8cm]{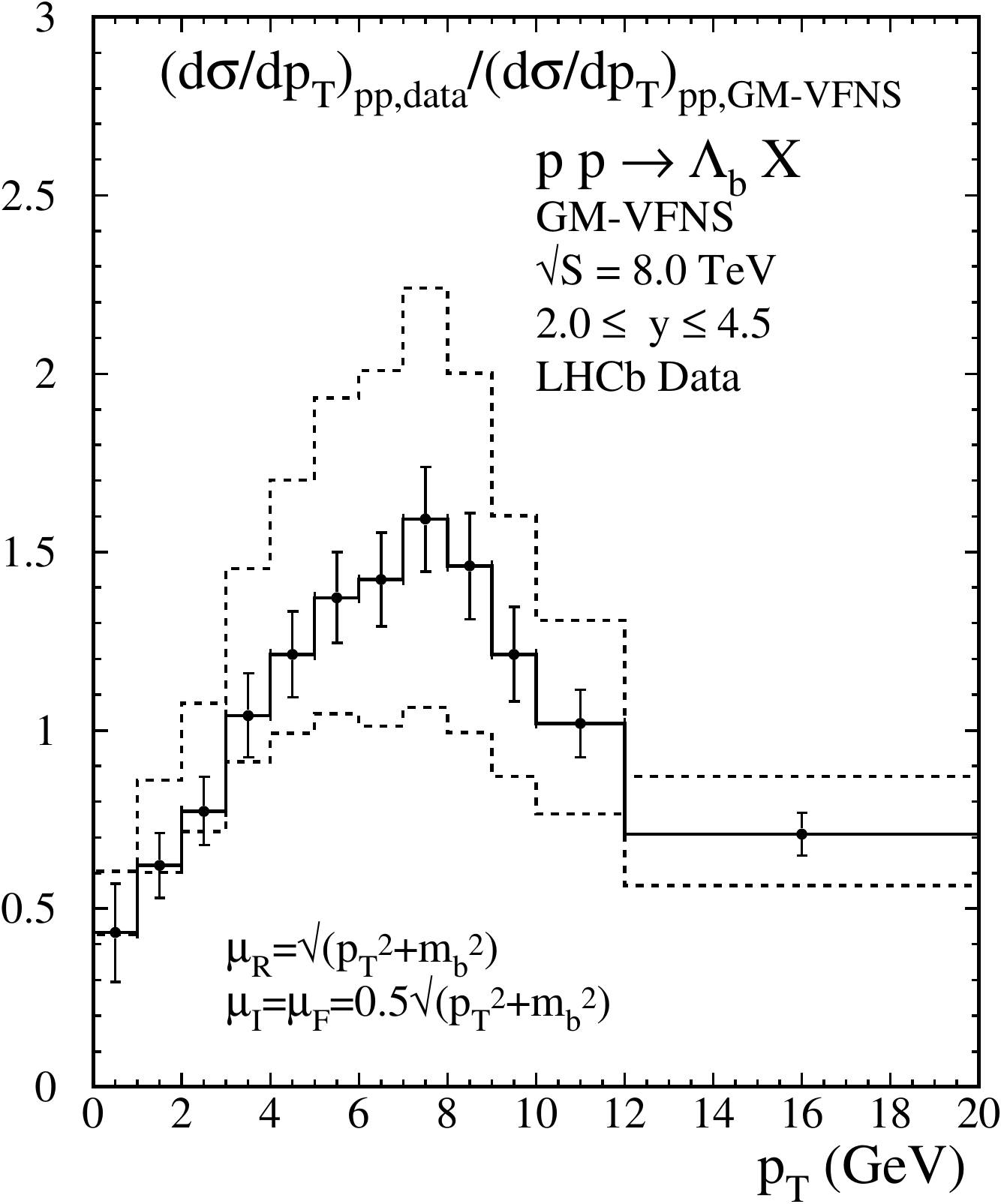}}
\end{center}
\caption{
  \label{fig:4} 
  Differential cross section $d\sigma/dp_T$ times branching 
  ratio for prompt inclusive $\Lambda_b^0$-baryon production 
  in the GM-VFNS for $\sqrt{S} = 8.0$~TeV $pp$ collisions with 
  $2.0 \leq y \leq 4.5$ compared to LHCb data~\cite{Aaij:2015fea}. 
  The upper and lower dashed histograms are calculated with 
  $\mu_R$ changed by factors $1/2$ and $2$. The ratios of data 
  over theory in the right panel are calculated as described 
  above (see caption of Fig.~\ref{fig:1}).
  }
\end{figure}

Our results are shown in Fig.~\ref{fig:3} for $\sqrt{S} = 
7$~TeV and in Fig.~\ref{fig:4} for $\sqrt{S} = 8$~TeV. 
Full-line histograms show the results for the default 
scales $\mu_F = 0.5 \sqrt{m_b^2 + p_T^2}$, $\mu_R = 
\sqrt{m_b^2 + p_T^2}$; the dashed-line histograms represent 
the estimate of theoretical uncertainties due to the 
variation of the renormalization scale. In the right panels 
of these figures we display the ratios of data over theory. 
As above we show the experimental error bars only for the 
central prediction, but the maximal and minimal values of 
the ratios have errors of the same magnitude as the central 
prediction. Taking account of these experimental uncertainties, 
as well as of uncertainties due to scale variations, we 
find in general a good agreement between data and theory, 
both for $\sqrt{S} = 7$~TeV and for $\sqrt{S} = 8$~TeV. 
The exceptions are the two data points at the lowest $p_T$ 
values and maybe the one at largest $p_T$. 

\begin{figure}[tb]
\begin{center}
\includegraphics[width=7.8cm]{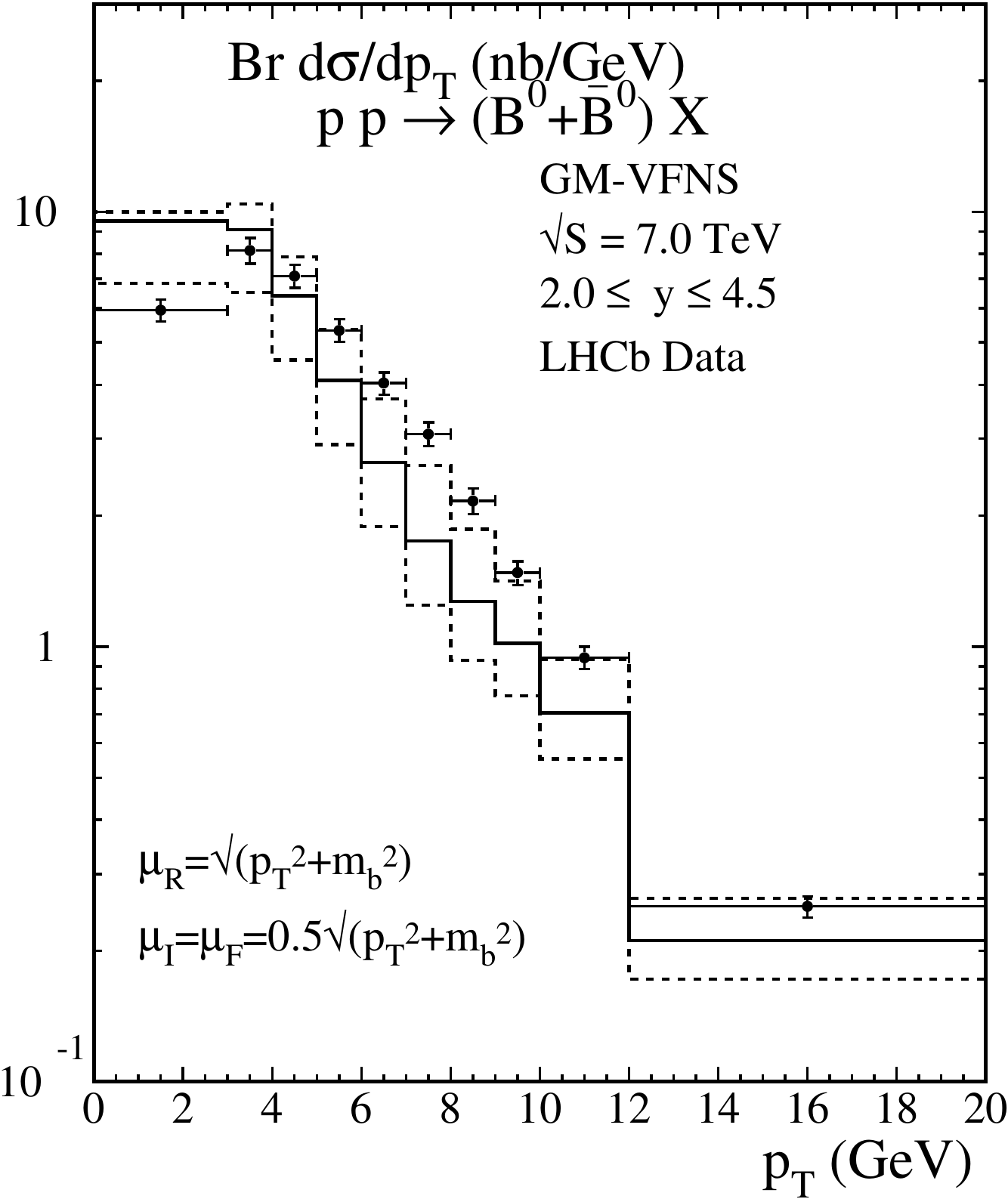}
\includegraphics[width=7.8cm]{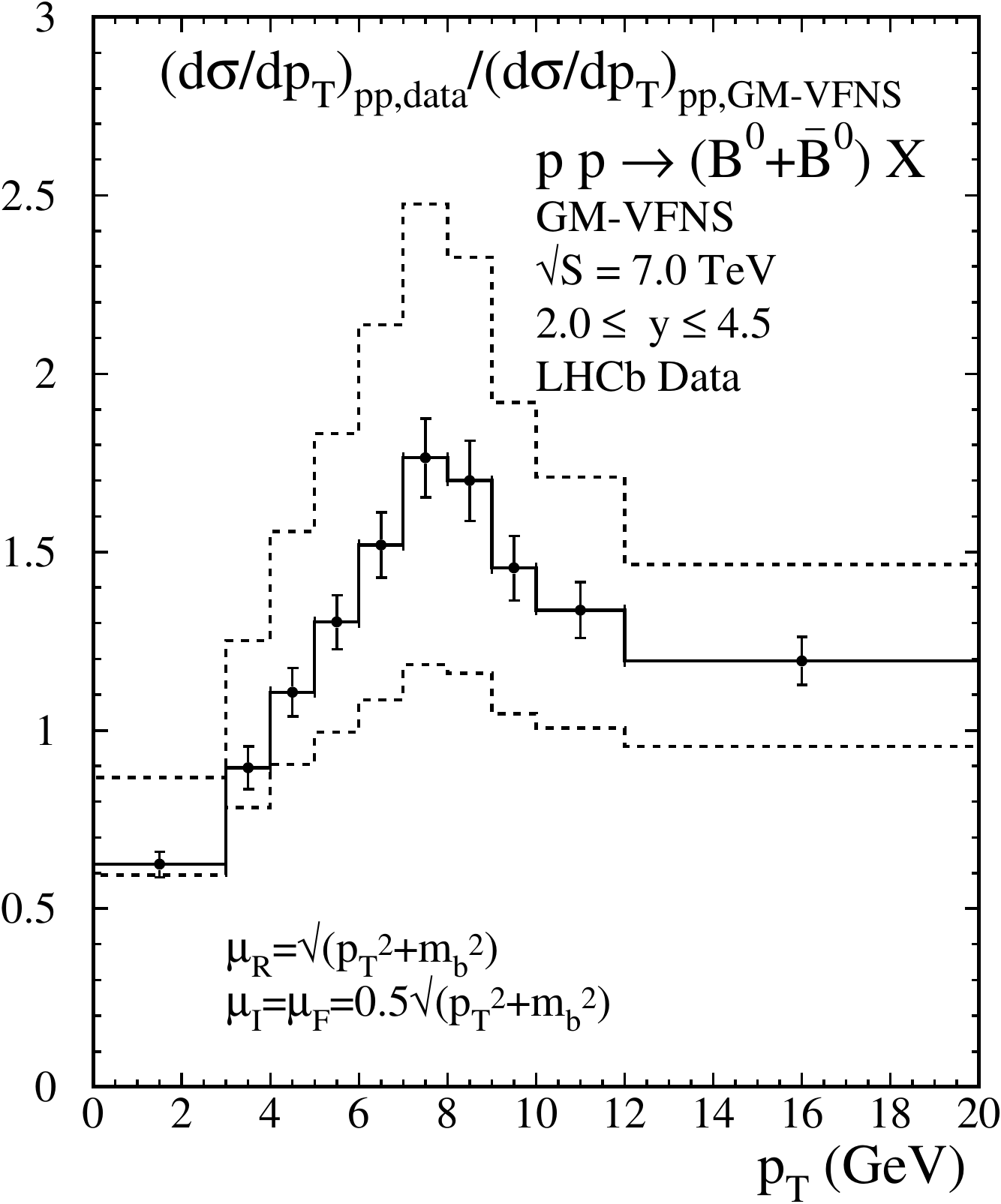}
\end{center}
\caption{
  \label{fig:5} 
  Differential cross section $d\sigma/dp_T$ times branching 
  ratio of prompt inclusive $(B^0+\bar{B}^0)$-meson production 
  in the GM-VFNS for $\sqrt{S} = 7.0$~TeV $pp$ collisions with 
  $2.0 \leq y \leq 4.5$ compared to LHCb data~\cite{Aaij:2015fea}. 
  The upper and lower dashed histograms are calculated with $\mu_R$ 
  changed by factors $1/2$ and $2$. The ratios of data over theory 
  in the right panel are calculated as described above (see caption 
  of Fig.~\ref{fig:1}).
  }
\end{figure}

\begin{figure}[tb]
\begin{center}
\includegraphics[width=7.8cm]{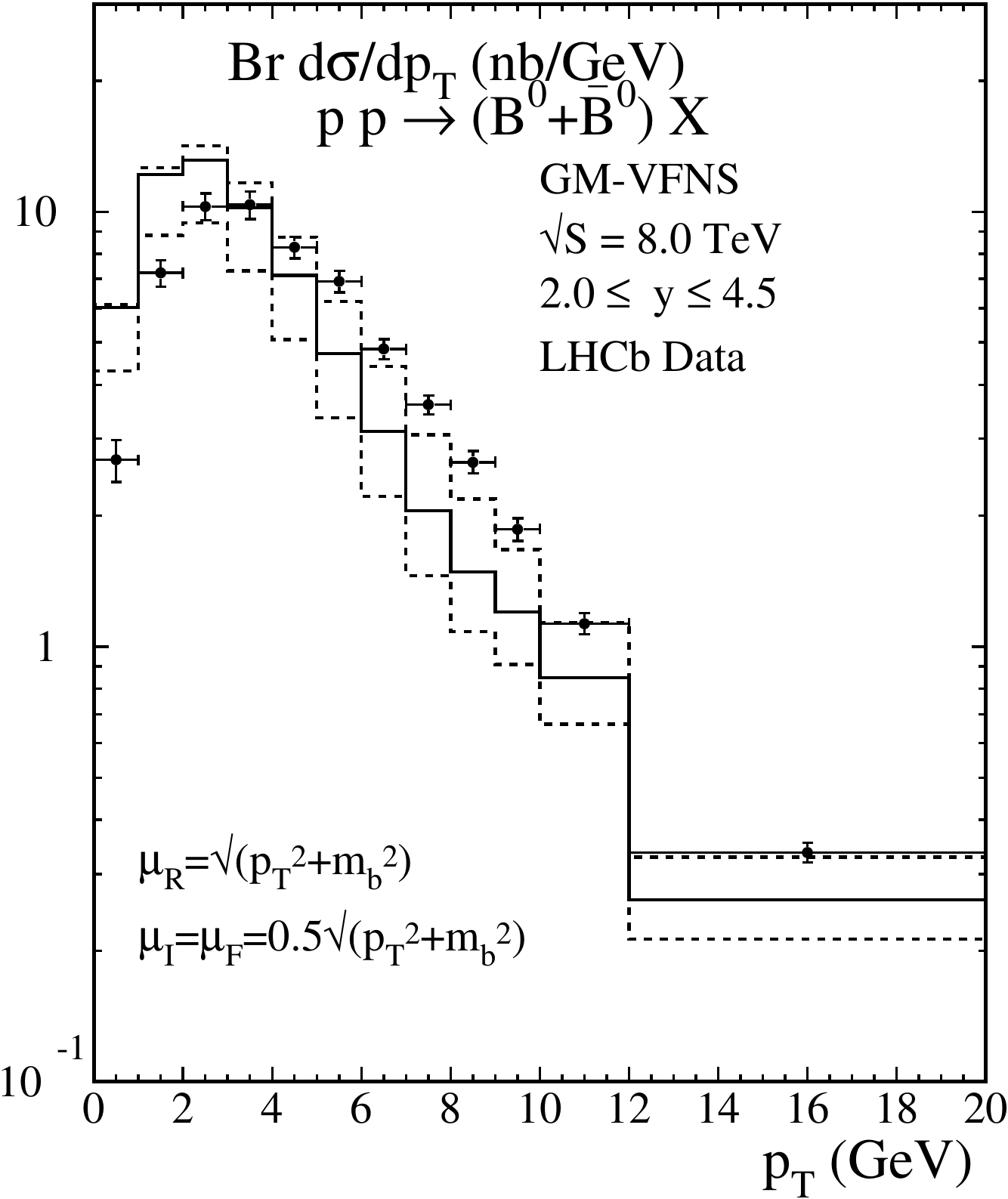}
\includegraphics[width=7.8cm]{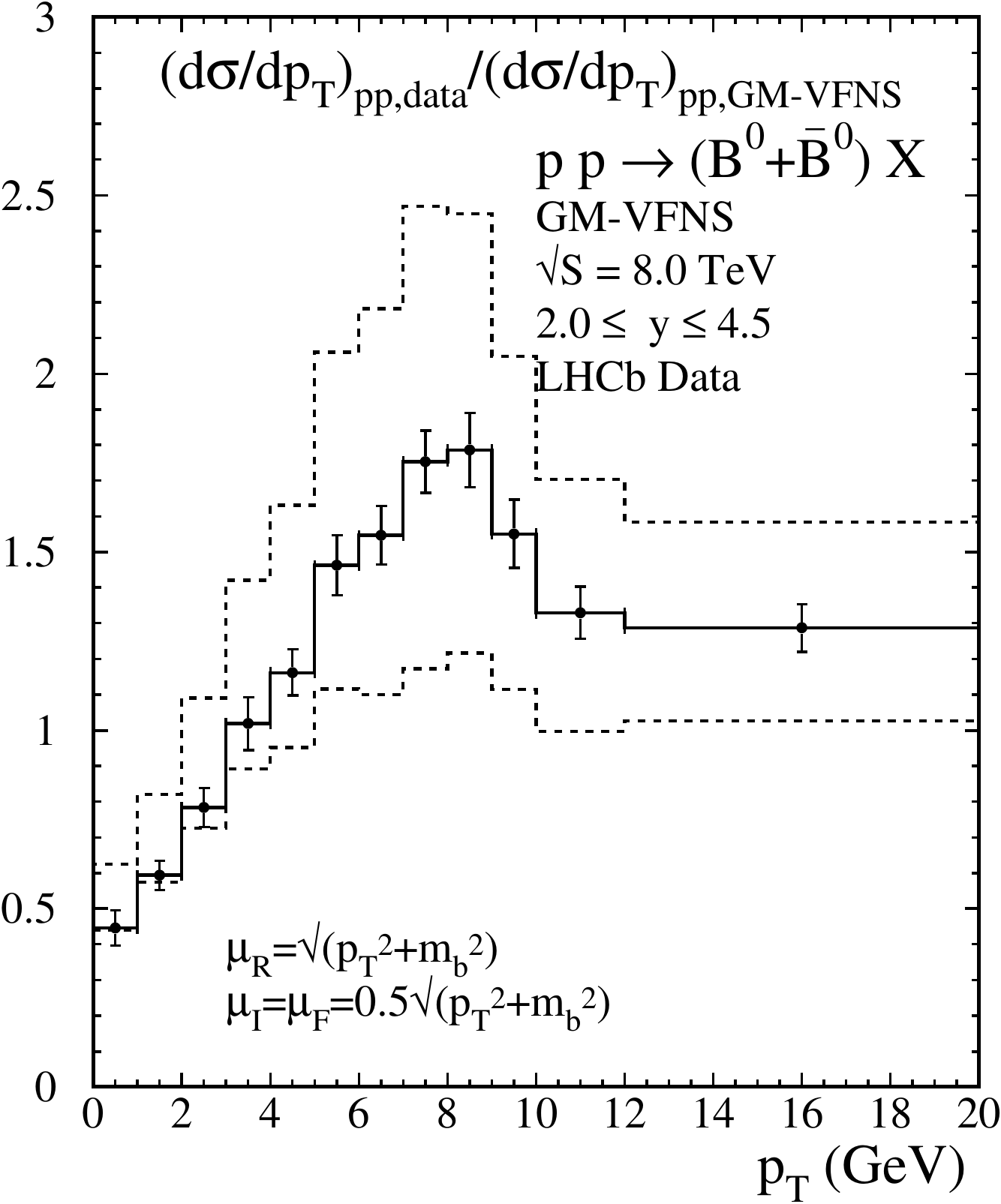}
\end{center}
\caption{
  \label{fig:6} 
  Differential cross section $d\sigma/dp_T$ times branching ratio 
  of prompt inclusive $(B^0+\bar{B}^0)$-meson production in the 
  GM-VFNS for $\sqrt{S} = 8.0$~TeV $pp$ collisions with $2.0 \leq 
  y \leq 4.5$ compared to LHCb data~\cite{Aaij:2015fea}. The upper 
  and lower dashed histograms are calculated with $\mu_R$ changed 
  by factors $1/2$ and $2$. The ratios of data over theory in 
  the right panel are calculated as described above (see caption 
  of Fig.~\ref{fig:1}).
  }
\end{figure}

We should remember that the FFs have been determined from 
$e^+ e^-$ data which are dominated by $B$-meson production. 
It is therefore instructive to verify that they can also 
be used to describe $B$-meson production in $pp$ collisions. 
For this purpose we show plots for the sum of $B^0 + \bar{B}^0$ 
production at 7 and 8~TeV in Figs.~\ref{fig:5} and \ref{fig:6}, 
respectively. As before, we show both the differential cross 
sections $d\sigma/dp_T$ times branching ratio (left panels) 
and ratios of data over theory (right panels) and compare with 
LHCb data~\cite{Aaij:2015fea}. The agreement of the data with 
our predictions looks very similar to the case of $\Lambda_b^0$ 
production, except for the data point at the largest $p_T$: 
for $B^0+\bar{B}^0$ there is good agreement within errors, 
whereas for $\Lambda_b^0$ production the corresponding data 
point was outside the error bands. This could indicate that 
the $\Lambda_b^0$-production cross section decreases somewhat 
faster with increasing $p_T$ than the $B^0+\bar{B}^0$-production 
cross section. 

\begin{figure}[tb]
\begin{center}
\includegraphics[width=7.8cm]{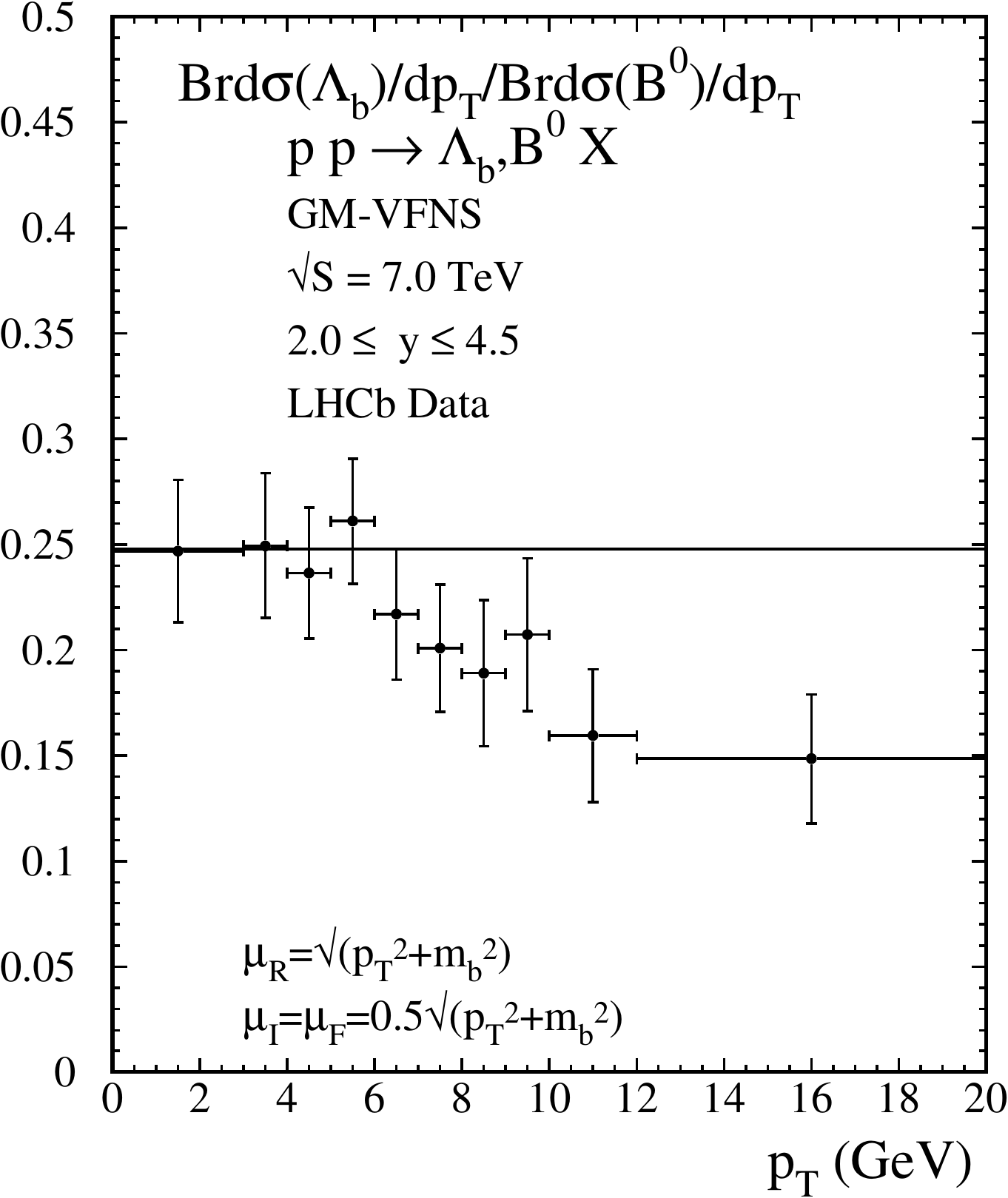}
\includegraphics[width=7.8cm]{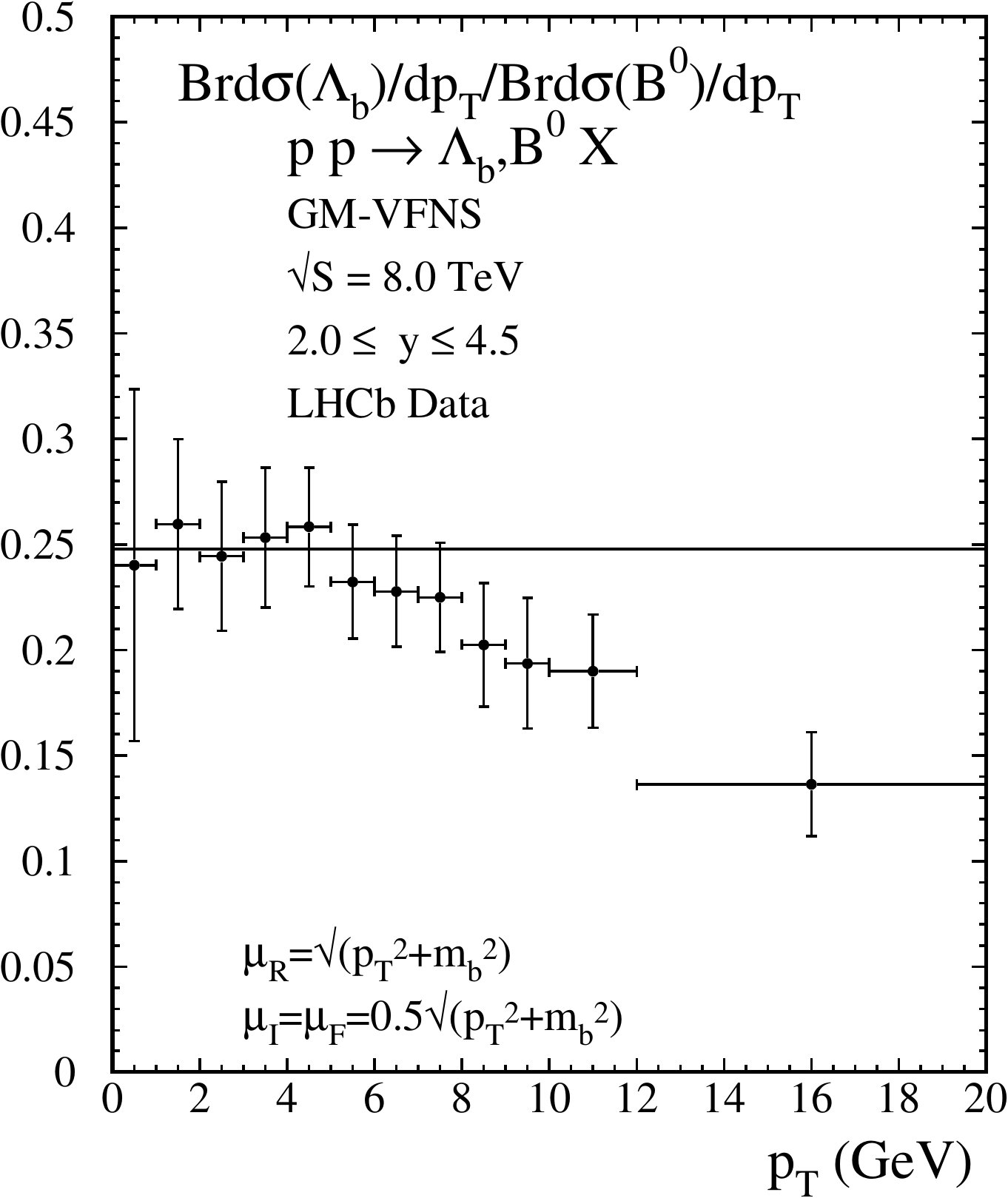}
\end{center}
\caption{
  \label{fig:7} 
  Ratios of production cross sections $d\sigma/dp_T$ times 
  branching fractions for prompt inclusive $\Lambda_b^0$ 
  over $B^0$ production as a function of $p_T$ for $\sqrt{S} 
  = 7.0$~TeV (left) and $\sqrt{S} = 8.0$~TeV (right).
  } 
\end{figure}

Differences between $B$-meson and $\Lambda_b$-baryon production 
are more clearly exhibited in Fig.~\ref{fig:7}. Here we show 
the ratios of the $p_T$ distributions for inclusive $\Lambda_b^0$ 
over $B^0$ production as a function of $p_T$. The differential 
cross sections $d\sigma/dp_T$ are multiplied with the respective 
branching fractions to obtain the results shown in this figure 
for $\sqrt{S} = 7$~TeV (left panel) and $\sqrt{S} = 8$~TeV (right 
panel). The horizontal line at the ratio $=0.248$ represents 
the theoretical prediction. It is independent 
of $p_T$ since we have assumed that the same FF is responsible 
for $b$-meson and for $b$-hadron production, i.e., only the 
corresponding branching fraction times the ratio of the 
fragmentation fractions $f_{\Lambda_b}/f_d$ has to be 
calculated. The experimental values for this ratio decrease 
with increasing $p_T$, fall below the theoretical prediction 
above $p_T = 7$~GeV (8 GeV) for $\sqrt{S} = 7$~ TeV (8~TeV), 
respectively, and reach the value $\simeq$ 0.15 in the bin 
with largest $p_T$. 

We emphasize that this ratio is particularly well suited to 
compare the FFs for $b \to \Lambda_b^0$ and $b \to B^0$ 
fragmentation since it is not affected by theoretical 
uncertainties due to scale variations. The errors shown 
in Fig.~\ref{fig:7} are purely experimental and we have added 
the uncertainties of the corresponding cross sections in 
quadrature. A full experimental analysis as performed in 
Ref.~\cite{Aaij:2015fea} can take into account correlations 
which leads to a partial cancellation of uncertainties in 
the ratio. Results\footnote{
  Note, however, that we have defined the $\Lambda_b^0 / B^0$ 
  ratio as normalized to the cross section for $B^0$ 
  production, not for $B^0 + \bar{B}^0$ production as in 
  Ref.~\cite{Aaij:2015fea}. 
}
have been shown in Fig.~6 of Ref.~\cite{Aaij:2015fea} with a 
similar conclusion, but significantly smaller errors.

\begin{figure}[tb]
\begin{center}
\includegraphics[width=7.8cm]{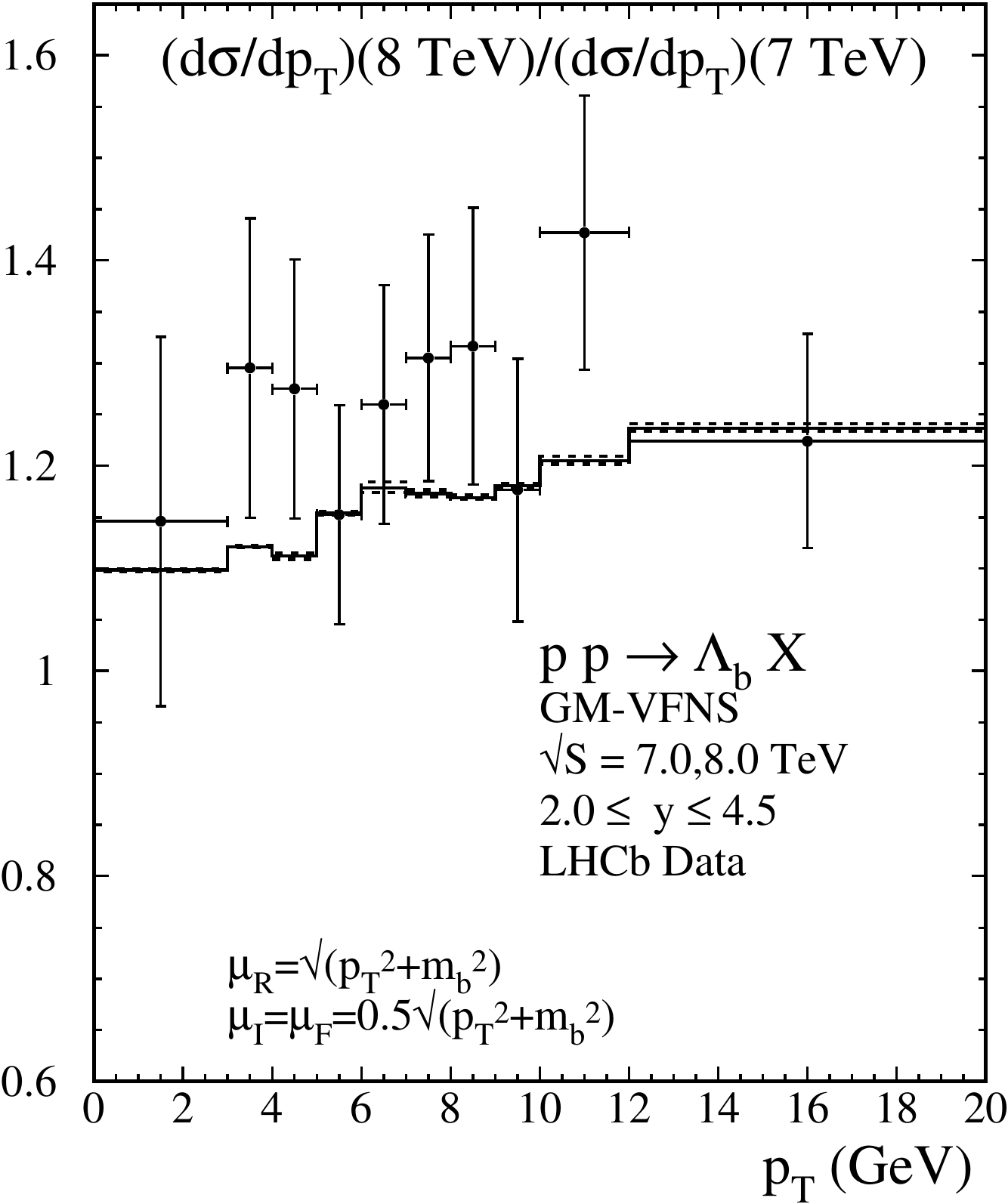}
\raisebox{-0.4mm}{\includegraphics[width=7.8cm]{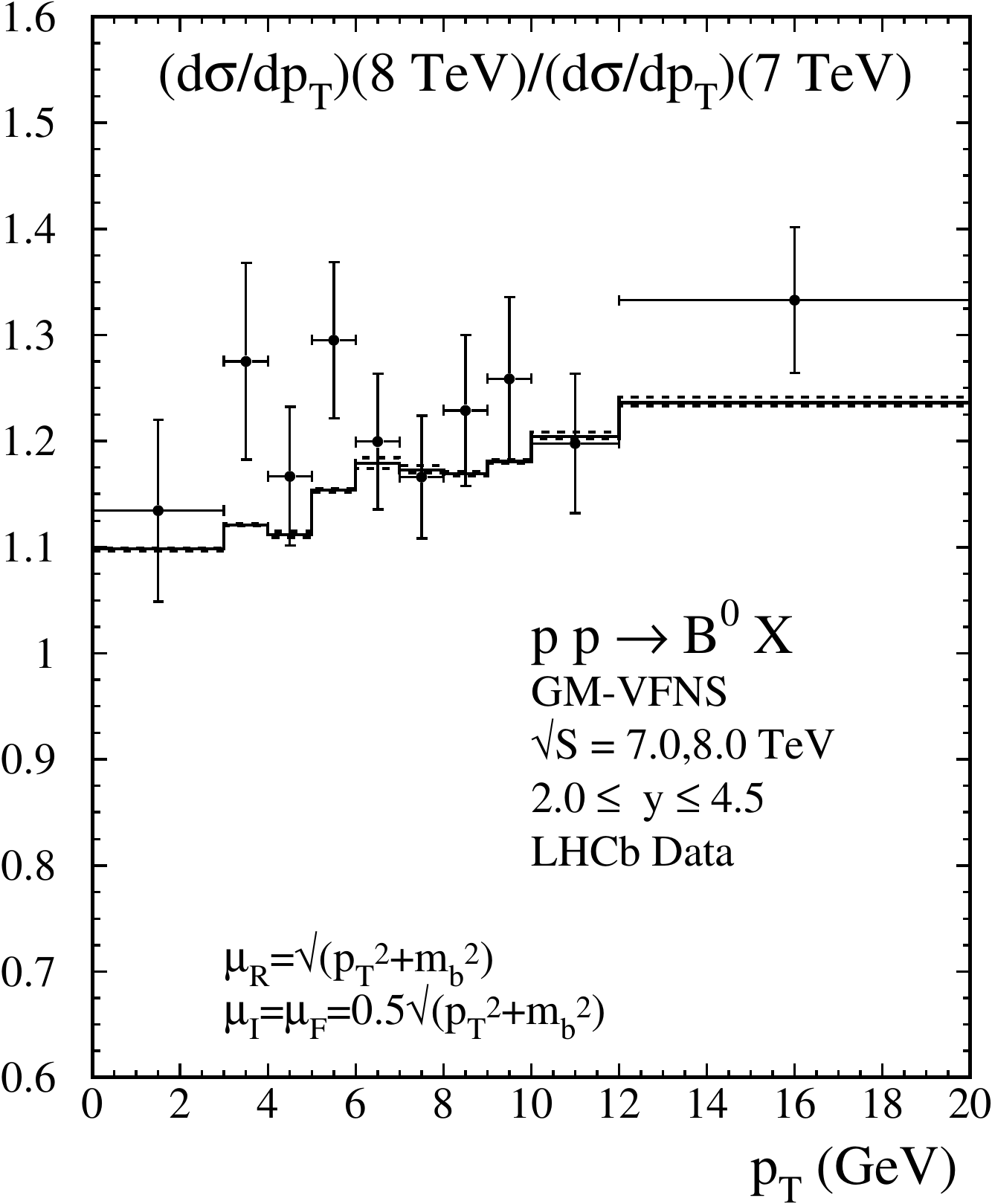}}
\end{center}
\caption{
  \label{fig:8} 
  Cross section ratios $d\sigma/dp_T$ for prompt $\Lambda_b^0$ 
  (left) and $B^0$ (right) production at $\sqrt{S} = 8.0$~TeV 
  over $\sqrt{S} = 7.0$~TeV compared with LHCb 
  data~\cite{Aaij:2015fea}.
  } 
\end{figure}

It may also be instructive to study the $\sqrt{S}$ dependence 
of the production cross sections. In Fig.~\ref{fig:8} we show 
the ratios of cross sections at $\sqrt{S} = 8$~TeV over 
those at $\sqrt{S} = 7$~TeV for both $\Lambda_b^0$ (left 
panel) and $B^0$ production (right panel). Our calculation 
is compared with LHCb data \cite{Aaij:2015fea}. For both final 
states the ratios vary between 1.1 and 1.2 as a function of 
$p_T$. For $B^0$ production the agreement with data is somewhat 
better than for $\Lambda_b^0$ production, but the uncertainties 
and statistical fluctuations are still too large to draw a 
definitive conclusion. 

\begin{figure}[tb]
\begin{center}
\includegraphics[width=7.8cm]{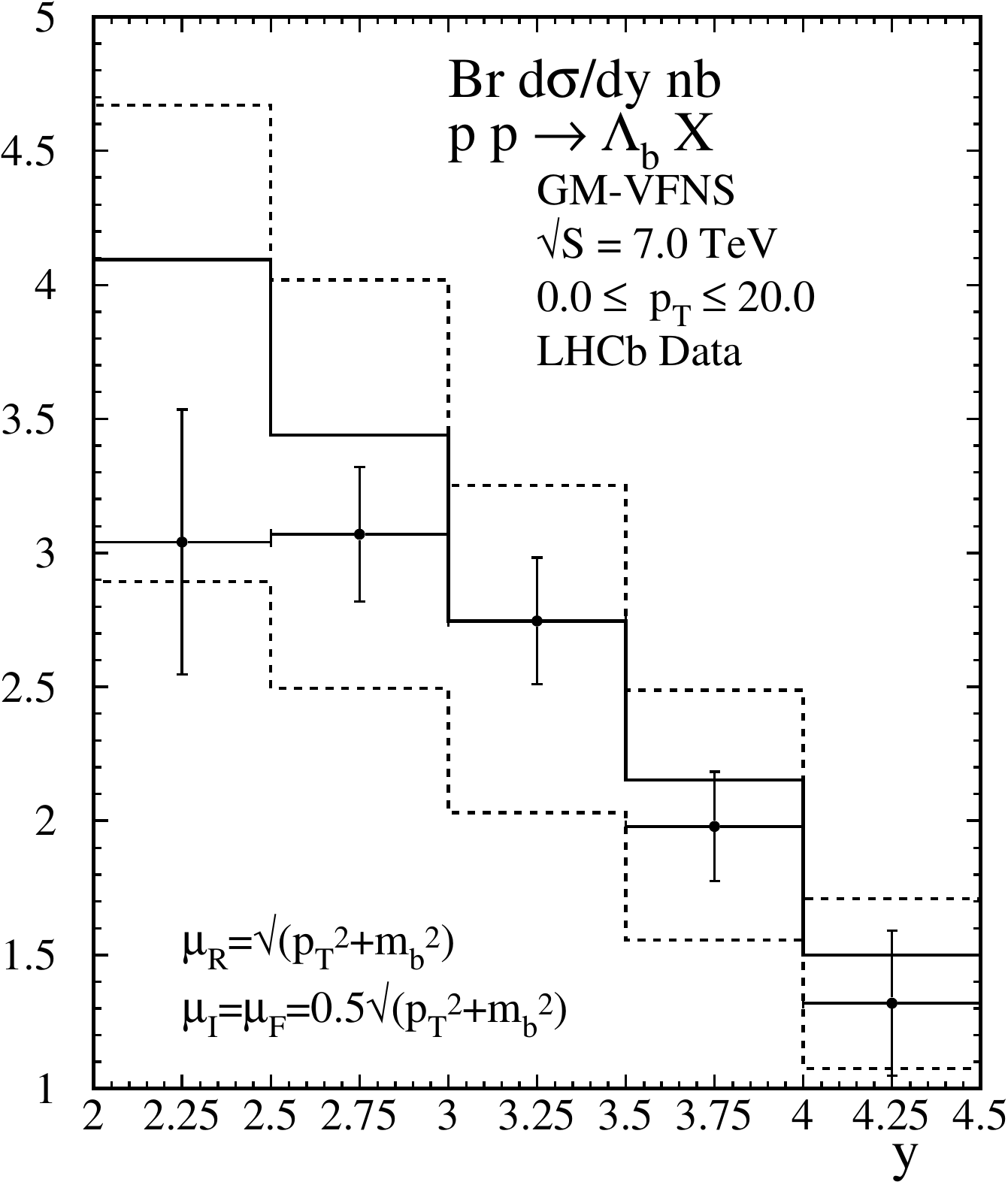}
\raisebox{0.4mm}{\includegraphics[width=7.8cm]{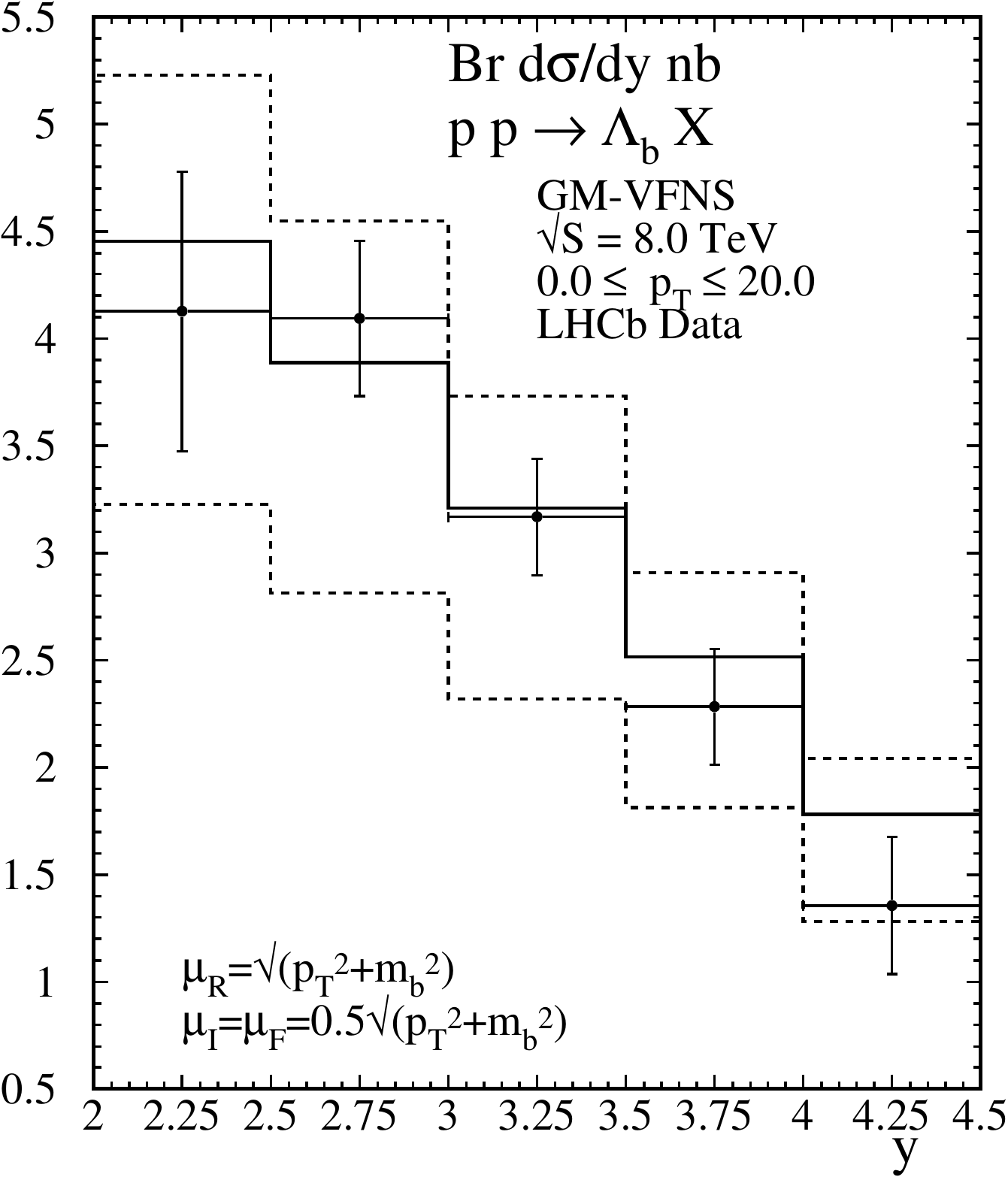}}
\end{center}
\caption{
  \label{fig:9} 
  Cross sections $d\sigma/dy$ times branching ratio for 
  prompt inclusive $\Lambda_b^0$ production at $\sqrt{S}=7.0$ 
  TeV (left) and $\sqrt{s}=8.0$ (right) compared to LHCb data 
  \cite{Aaij:2015fea}.
  }
\end{figure} 

\begin{figure}[tb]
\begin{center}
\includegraphics[width=7.8cm]{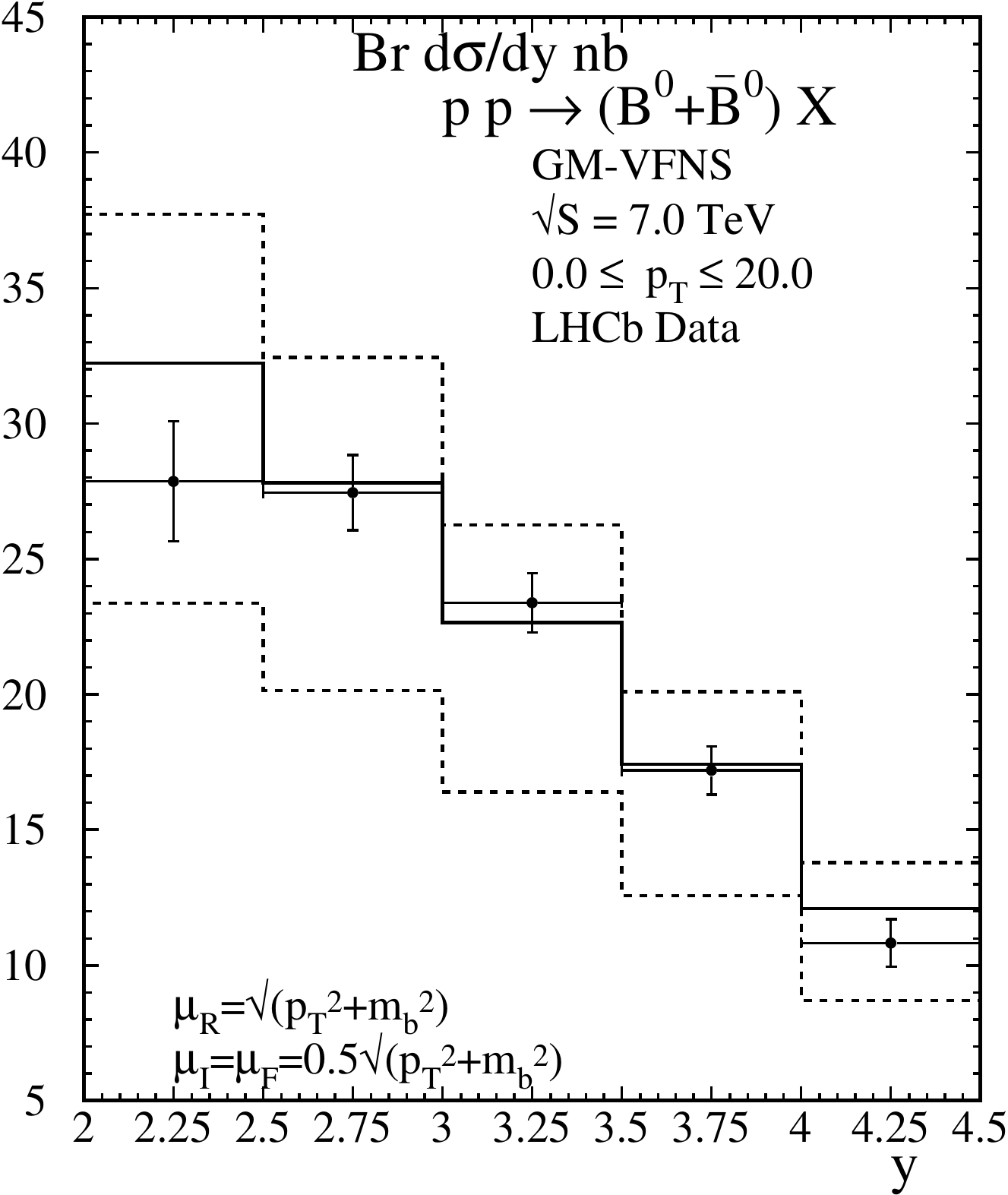}
\includegraphics[width=7.8cm]{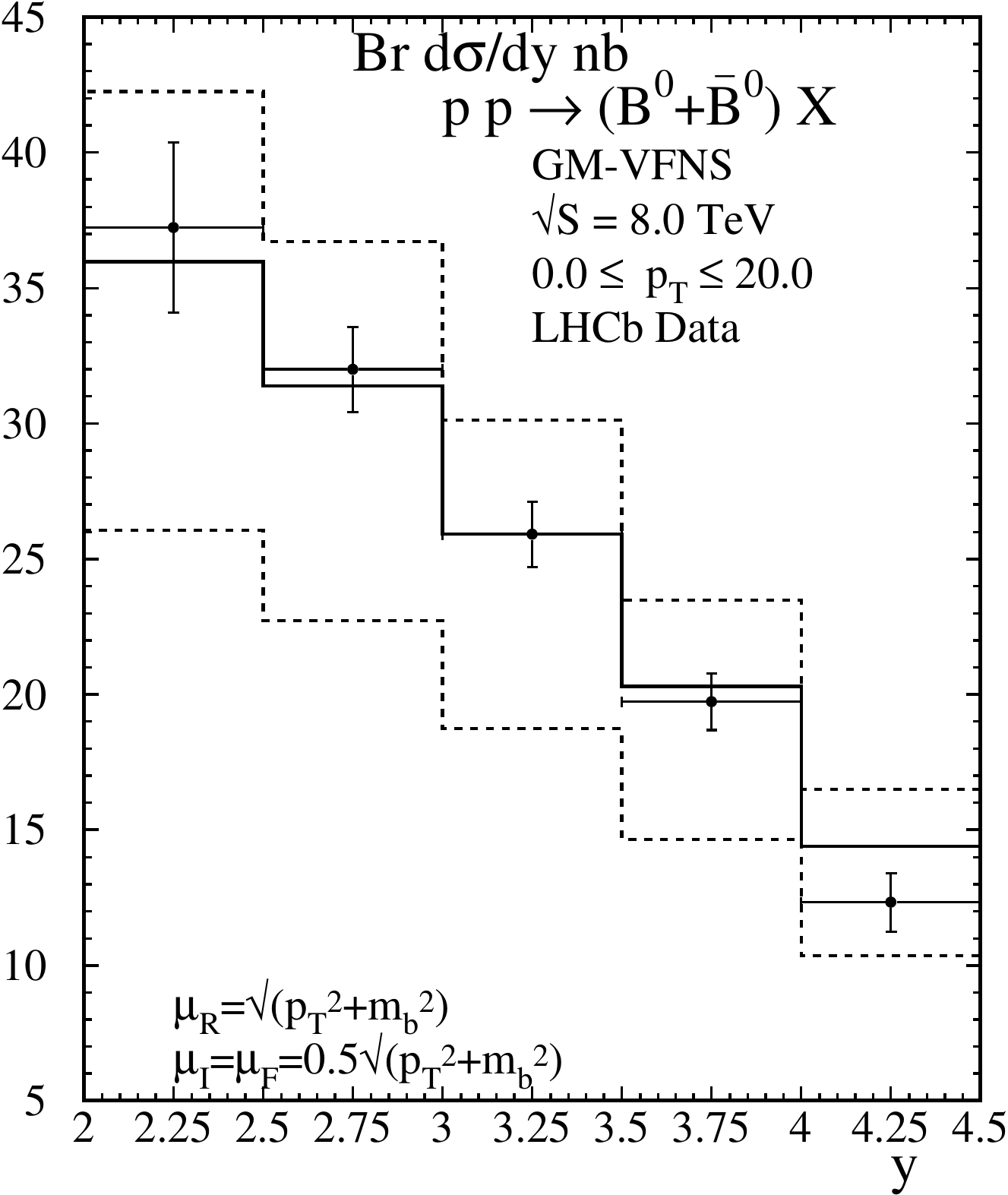}
\end{center}
\caption{
  \label{fig:10} 
  Cross sections $d\sigma/dy$ times branching ratio for 
  prompt inclusive $B^0+\bar{B}^0$ production at $\sqrt{S}=7.0$ 
  TeV (left) and $\sqrt{s}=8.0$ (right) compared to LHCb data 
  \cite{Aaij:2015fea}.
  }
\end{figure} 

\begin{figure}[tb]
\begin{center}
\includegraphics[width=7.8cm]{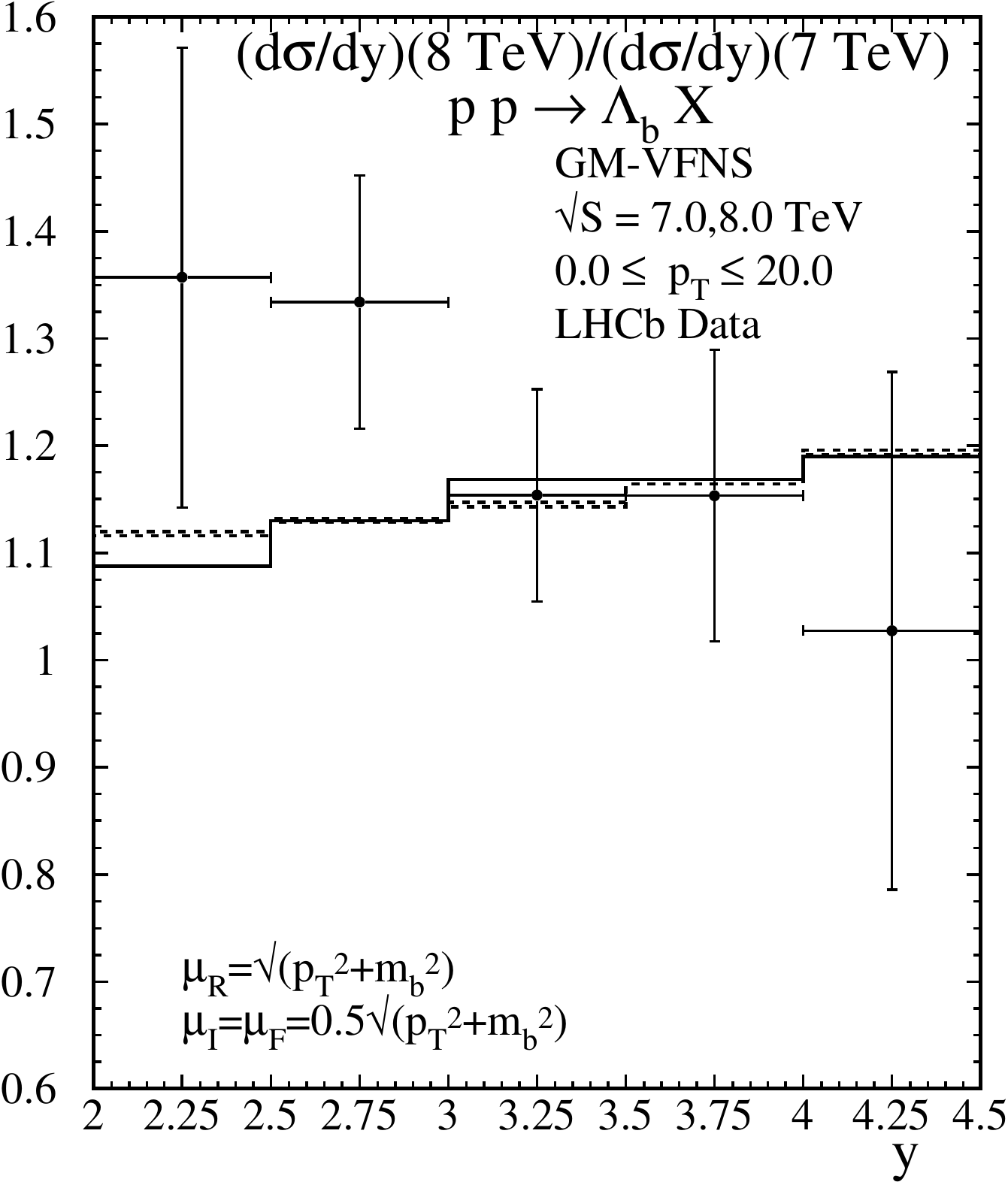}
\includegraphics[width=7.8cm]{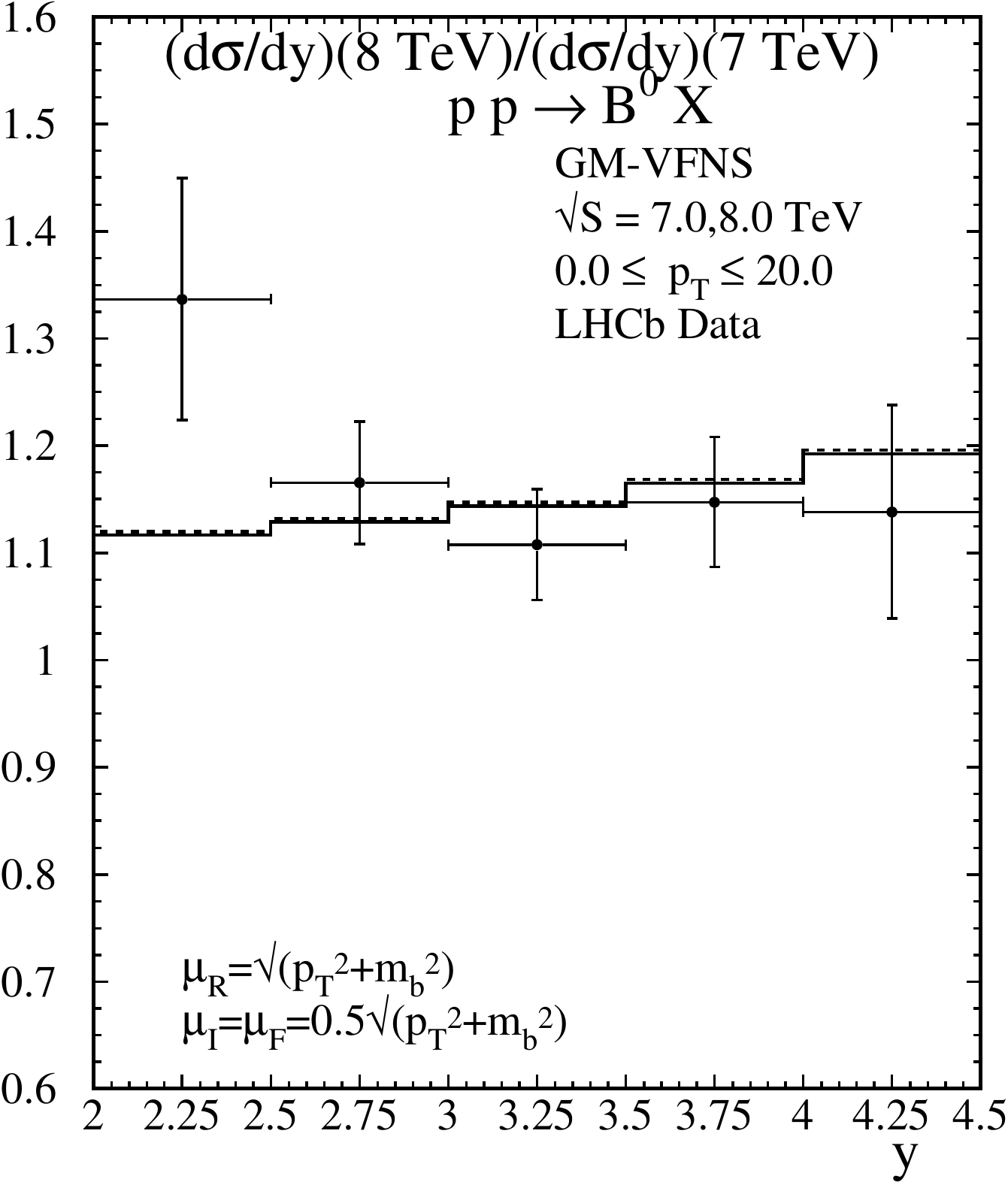}
\end{center}
\caption{
  \label{fig:11} 
  Ratio of production cross sections $d\sigma/dy$ for $\sqrt{s} 
  = 8.0$~TeV over $\sqrt{s} = 7.0$~TeV for $\Lambda_b^0$ (left) 
  and $B^0$ production (right) compared to LHCB 
  data~\cite{Aaij:2015fea}.
  } 
\end{figure} 

Finally we present results for the rapidity distributions. 
The cross sections $d\sigma/dy$ times the respective branching 
ratios for $\Lambda_b^0$ and $B^0$ production as a function 
of $y$ for five bins in the range $2.0 < y < 4.5$ are shown 
in Fig.~\ref{fig:9}. The left and right panels are for $\sqrt{S} 
= 7$ and 8~TeV, respectively. Our predictions are compared with 
LHCb data~\cite{Aaij:2015fea}. We have obtained the corresponding 
cross section values from tables 3 -- 6 of Ref.~\cite{Aaij:2015fea}, 
summing the data for the double-differential cross sections 
$d^2\sigma/dp_Tdy$ given there over the $p_T$ bins in the range 
$0 < p_T < 20$ GeV. For each $p_T$ bin, statistical and 
systematic errors are added linearly first, then the errors 
are combined quadratically to obtain the total uncertainties 
shown in Fig.~\ref{fig:9}. We find quite satisfactory agreement 
between predictions and data. Similar results for $B^0 + 
\bar{B}^0$ production are shown in Fig.~\ref{fig:10}, again 
for $\sqrt{S} = 7$ (left panel) and 8~TeV (right panel) and 
also compared with experimental data from Ref.~\cite{Aaij:2015fea}. 
The agreement between predictions and data is similarly good 
as for the case of $\Lambda_b^0$ production.

Ratios of $d\sigma/dy$ for $\sqrt{S} = 7$ over 8~TeV for 
$\Lambda_b^0$ and $B^0$ production have been shown in 
Ref.~\cite{Aaij:2015fea} as well. We present corresponding 
theoretical predictions in Fig.~\ref{fig:11}. Again, our 
calculation of the errors does not take into account 
correlations and the experimental uncertainties found in 
Ref.~\cite{Aaij:2015fea} are somewhat smaller than in our 
plots. Apart from this difference we find results which are 
quite similar to the ratios calculated with the FONNL 
approach \cite{Cacciari:1998it,Cacciari:2012ny} also given 
in Ref.~\cite{Aaij:2015fea}.


\section{Conclusions}

We have performed a detailed study of next-to-leading-order 
predictions for inclusive $b$-hadron production in $pp$ collisions 
within the general-mass variable-flavor-number scheme. Our 
predictions are based on the assumption that $B$-meson and 
$\Lambda_b^0$-baryon production can be described by a common 
fragmentation function and that only constant branching fractions 
have to be chosen appropriately. The comparison with data for 
$\Lambda_b^0$-baryon production from the CMS and the LHCb 
collaborations at the CERN LHC shows agreement in the overall 
picture. However, at larger transverse momenta, the data from both 
experiments, which cover different rapidities, fall below the 
predictions. In particular the ratio of $\Lambda_b^0$-baryon over 
$B$-meson production exhibits indications that the fragmentation 
functions need to be modified at larger values of the scale 
variable. We expect that future data with reduced experimental 
uncertainties will help to clarify the situation.


\section*{Acknowledgment}

We thank Michael Schmelling for helpful correspondence  
about the LHCb publication~\cite{Aaij:2015fea}.


\end{document}